\documentclass[twocolumn]{aastex701}
\usepackage{natbib}

\begin{document}

\title{Resonant Super-Earths Dancing With EKL Oscillations: TTV Phase Excitation and Resonance Disruption by EKL Interactions between a Cold Jupiter and Stellar Companion}
\author[]{Pin-Gao Gu}
\affiliation{Institute of Astronomy and Astrophysics, Academia Sinica, Taipei 106319, Taiwan}
\email[show]{gu@asiaa.sinica.edu.tw}  



\author{Gongjie Li}
\affiliation{Center for Relativistic Astrophysics, School of Physics, Georgia Institute of Technology, Atlanta, GA 30332, USA}
\email{fakeemail4@google.com}





\begin{abstract}
Near-resonant Kepler planets are dynamically hot, as evidenced by nonzero transit timing variation (TTV) phases, indicating that free eccentricities are not damped. Recent observations suggest that circulating near-resonant planets tend to be dynamically unstable, and hence dynamically hot, likely representing an intermediate stage in the close-in super-Earth population at young ages. We investigate whether a cold Jupiter interacting with a stellar companion through the eccentric Kozai-Lidov mechanism (EKL) can excite TTV phases and increase the libration amplitude of resonant angles in close-in resonant pairs.
We find that the EKL model that drives the observed eccentricity of cold Jupiters can also excite TTV phases, increase the libration amplitude of resonant angles away from ideal geometric alignment, and even disrupt them in a significant fraction of planetary systems in our simulated samples over 16 Myr. We also find that the TTV phases of the resonant pairs tend to be small ($< 90^\circ$), while the resonant angles are more easily elevated to become circulating during EKL excitation. 

\end{abstract}



\section{Introduction}
Transit timing variations (TTVs) are among the most incisive probes of the dynamical architecture and evolutionary history of compact multi-planet systems. Mutual gravitational perturbations among transiting planets cause deviations from strictly periodic transit times, enabling the inference of masses, eccentricities, and resonant states directly from photometry \citep[][and references therein]{2005MNRAS.359..567A,2005Sci...307.1288H,2017AJ....154....5H,2018haex.bookE...7A}. Near first-order mean-motion resonances (MMRs), the dominant timing signal varies on a resonant super-period set by proximity to commensurability. In this regime, analytic theory shows that the TTV amplitude is sensitive to both masses and eccentricities, whereas the TTV phase, the phase offset of the timing signal relative to conjunction at a transit, depends strongly on the planets' free eccentricities, i.e., those not fixed by resonant forcing.  A key lesson from TTV measurements is that non-zero phase offsets are common in near-resonant systems. Without free eccentricity, resonant TTVs align in phase with the forcing terms in ideal geometric alignment; non-zero observed phase shifts instead require free eccentricities comparable to or exceeding the forced component \citep{2012ApJ...761..122L,2012ApJ...756L..11L}.

Signatures of resonance and near-resonance are also imprinted in planetary architecture. Most of the Kepler super-Earths, which are $\sim$ Gyr old, are not near-MMRs \citep[e.g.,][]{2014ApJ...790..146F}. In contrast, recent Kepler-Gaia analyses and small-statistical surveys suggested that near-resonant super-Earth systems are more frequent among young multi-planet systems, with near-commensurable period ratios declining with stellar age, consistent with resonant chains forming early and later evolving toward non-resonant architectures \citep[e.g.,][]{2024AJ....168..239D,2024AJ....167...55H,2025ApJ...995..206H}. This age dependence underscores that non-vanishing free eccentricities and non-equilibrium resonant angles are not rare; rather, they reflect an intermediate dynamical stage in the system's evolution. Numerical simulations showed that near-resonant super-Earths are dynamically fragile, consistent with these observational suggestions \citep{2025AJ....169..323L}.

Such non-zero TTV phases imply dynamically excited configurations, which are at odds with classic formation scenarios that involve smooth, strongly damped Type-I migration into low-eccentricity resonant equilibria in a protoplanetary disk \citep[e.g.,][]{Tanaka_2002}. Non-zero TTV phases typically correlate with inferred large resonant angles, indicating that many systems either librate with high amplitude or circulate, rather than sit near equilibrium resonant fixed points \citep[e.g.,][]{2023ApJ...948...12G}. Recent theoretical work has sharpened the connection between TTV phases and dynamical histories. 
Displaced resonant states can arise if Type-I migration in a laminar disk is no longer valid for super-Earths \citep[e.g.,][]{2021ApJ...921..142C}, if resonant capture is stochastic in a turbulent disk \citep[e.g.,][]{2008ApJ...683.1117A,2025MNRAS.540.1998C}, or if subsequent dynamical perturbations excite eccentricities after the natal disk disperses, such as divergent migrations through scattering by small bodies, gravitational interactions with the escaping atmosphere, secular interactions with outer eccentric embryos, or gravitational scattering with planetesimal flybys \citep[e.g.,][]{Chatterjee_2015,2020ApJ...893...43M,2025ApJ...994..123L,2024ApJ...971....5W,2026arXiv260221349H,2025AJ....169...19H,2026ApJ...996...91O,2026ApJ...998L...5L}.

In particular,
\citet{2023MNRAS.522.1914C} showed that TTV phase offsets in a resonant pair can be excited by secular forcing from a coplanar external companion. This model is more receptive to observations, as external giant planets are more readily detectable and can be used to test it. Moreover, secular perturbations from an inclined external giant planet (cold Jupiter) were proposed to excite mutual inclinations in multiple planetary systems and thus cause the Kepler dichotomy of super-Earths \citep[e.g.,][]{2017AJ....153...42L}. 
Furthermore, in the context of exoplanet demographics, several survey studies have suggested that the population of close-in super-Earths is not independent of the cold Jupiter population and may be correlated in their occurrence rates \citep[e.g.,][]{2018AJ....156...92Z,2025A&A...700A.126B,2025ApJ...982L...7B}. 

Cold Jupiters are theoretically expected to form in protoplanetary disks on low-eccentricity orbits \citep[][for a review]{2023ASPC..534..685P}. Observationally, however, the eccentricities of these planets at system ages of several Gyr are frequently found to be substantially elevated \citep[e.g.,][]{2024ApJ...962L..21K,2026ApJ...999L..26B}.
A variety of dynamical processes have been proposed to account for this eccentricity excess. Post-formation mechanisms include mutual gravitational scattering or secular chaos in systems hosting multiple cold Jupiters, as well as eccentricity oscillations induced by a distant stellar companion through the eccentric Kozai–Lidov (EKL) mechanism \citep[e.g.,][]{1996Sci...274..954R,2003ApJ...589..605W,2007ApJ...669.1298F,2016ARA&A..54..441N}. These evolutionary pathways operate on distinct characteristic timescales and can, under appropriate conditions, drive high-eccentricity tidal migration that produces hot Jupiters.

More recently, \citet{2025ApJ...980L..31W} assumed an initial Rayleigh distribution for the eccentricities of cold Jupiters and demonstrated that, when the mean eccentricity lies in the range 0.1–0.2, the resulting steady-state eccentricity distribution is consistent with that expected from EKL-driven evolution by stellar companions. Such an initial Rayleigh distribution may naturally arise from planet–planet scattering operating on Myr timescales following the dispersal of the protoplanetary disk.

The dynamical influence of eccentric, distant giant planets—whether their orbits are excited via planet–planet scattering or through stellar EKL perturbations—on close-in, low-mass planets has been studied extensively \citep[e.g.,][and references therein]{2017MNRAS.467.1531H,2017MNRAS.468.3000M,2017AJ....153..210H}. However, the specific ways in which these mechanisms affect the excitation of TTV phases and the libration characteristics of MMRs among young super-Earth systems remain largely unexplored, thereby deserving further detailed investigation. Very recently, \citet{2026ApJ..1003...45G} demonstrated that gravitational scattering between cold Jupiters can induce circulation of the resonant angles in a pair of close-in super-Earths.

Motivated by the recent studies mentioned above, we perform N-body simulations to interpret nonzero TTV phases and large libration of resonant angles, and, by extension, the dynamical activity in near-resonant systems under the secular influence of a cold Jupiter interacting with a stellar companion. In this work, we focus on resonant pairs rather than resonant chains to simplify the investigation of the dynamical excitation of a resonant system by stellar EKL at a clean and fundamental level. We follow \citet{2023MNRAS.522.1914C} in considering a librating resonant pair with an initial spacing wider than about 1\% of the exact MMRs. We shall apply the EKL oscillation model of \citet{2025ApJ...980L..31W} to an ensemble of planetary systems, each comprising a close-in super-Earth pair in near-2:1 or 3:2 MMRs, and study the resulting excitation of TTV phases and the resonance disruption of these near-resonant pairs.

\section{Numerical Setup}
\label{sec:setup}
We consider a pair of close-in super-Earths, denoted by SE1 (interior) and SE2 (exterior), around the host star of $m_*=1 M_\odot$.  The mass of each super-Earth is 8 $M_\oplus$, similar to that considered in the representative case of the TTV phase study by  \citet{2023MNRAS.522.1914C}.
Following the convergent migration prescription adopted by \citet{2022AJ....163..201G}, we set up an initial coplanar resonant pair of super-Earths just wide of period commensurabilities of $(q+1)/q=$ 2:1 and 3:2. This is achieved by applying dissipative forces in the \textsc{REBOUNDx} N-body code to damp the pair's semimajor axis and eccentricity until they are captured in near-MMRs at approximately 0.1 au \citep{2020MNRAS.491.2885T}. The eccentricity damping timescale must be sufficiently small to increase the final period ratios $P_{\rm SE2}/P_{\rm SE1}$ up to $\Delta \equiv [q/(q+1)](P_{\rm SE2}/P_{\rm SE1})-1 \sim 1$\% greater than the exact resonant integer $(q+1)/q$, an effect similar to resonant repulsion \citep[e.g.,][]{2012ApJ...756L..11L,2023AJ....165...33D}. 

Once the pair is captured in resonance, we turn off the dissipative forces and continue evolving the system for a while. At this point, for the 2:1 pair, their semimajor axes [$a_{\rm SE1}$, $a_{\rm SE2}$] $\approx$ [0.08, 0.127] au, eccentricities [$e_{\rm SE1}$, $e_{\rm SE2}$] $\approx$ [$1.8\times10^{-3}$, $5\times 10^{-4}$], $\Delta \approx 1$\%, and the libration amplitudes of resonant angles are $\Delta \phi_{1} \approx 12^\circ$ and $\Delta \phi_{2} \approx 40^\circ$(see the left panels of Figure \ref{fig:comp_init} in Appendix \ref{sec:capture}). The pair's eccentricities are dominated by the forced components. Specifically, the free and forced eccentricities of the inner (outer) planet are $\approx 1.8\times 10^{-4}$ and $1.8\times 10^{-3}$ ($\approx 1.0\times 10^{-4}$ and $5.2\times 10^{-4}$), respectively.
For the 3:2 pair, [$a_{\rm SE1}$, $a_{\rm SE2}$] $\approx$ [0.0957, 0.127] au, [$e_{\rm SE1}$, $e_{\rm SE2}$] $\approx$ [$7.5\times10^{-4}$, $8.0\times 10^{-4}$], $\Delta \approx 2.35$\%, and the libration amplitudes are $\Delta \phi_{1} \approx \Delta \phi_{2} \approx 40^\circ$. 
The eccentricities of the 3:2 pair are also dominated by the forced eccentricities; the free and forced eccentricities of the inner (outer) planet are $\approx 1.5\times 10^{-4}$ and $7.9\times 10^{-4}$ ($\approx 1.5\times 10^{-4}$ and $8.4\times 10^{-4}$), respectively.
The periapses of the 2:1 and 3:2 pairs librate near anti-alignment, i.e., $\Delta \varpi = \varpi_2-\varpi_1$ librates around 180$^\circ$.

We then add a coplanar cold Jupiter (CJ) and a stellar companion (star2) on an inclined orbit, with initial conditions following the ``r0.13" model in \citet{2025ApJ...980L..31W}, i.e., the initial eccentricity of the cold Jupiter, $e_{\rm CJ}$ is drawn from a Rayleigh distribution with a mean of 0.13. This is an EKL model consistent with the observed eccentricity distribution of cold Jupiters. Here, we summarize the other initial parameters in the r0.13 model and refer readers to \citet{2025ApJ...980L..31W} for references and a detailed description. For the cold Jupiter, its mass, $m_{\rm CJ}$, is drawn from a power-law distribution $dn/dm \propto m^{-1}$ in the range of 0.3 to 10 $M_{\rm Jup}$, and its initial semimajor axis, $a_{\rm CJ}$, is uniformly sampled from 0.5 to 6 au.  For the stellar companion, its initial eccentricity, $e_{\rm star2}$, is sampled uniformly from 0 to 1, its semimajor axis, $a_{\rm star2}$, is drawn from a lognormal distribution from 50 to 1500 au, and the cosine of its initial orbital inclination, $i_{\rm star2}$, is uniformly sampled from $-1$ to 1. The mean longitude $\lambda$ and the longitude of periapsis $\varpi$ of the cold Jupiter and the stellar companion are sampled uniformly from $0^\circ$ to $360^\circ$.  We consider a one-solar-mass companion for the fiducial case of this study. We note that our results do not change qualitatively compared to the case following a Salpeter power-law distribution for stellar masses between 0.6 and 1.6 solar masses (see section \ref{sec:stellar_mass}). In addition, we note that close separations of the stellar companions could lead to truncation of the protoplanetary disk, making it challenging to form the CJs via core accretion (see more discussions in section \ref{sec:CJformation} and Appendix \ref{sec:truncation}). We use the IAS15 integrator of \textsc{Rebound} for all simulations \citep{2012A&A...537A.128R,2015MNRAS.446.1424R}.

Unlike \citet{2023MNRAS.522.1914C}, we exclude CJs during the capture of the SEs into MMRs. Since an initial Rayleigh distribution of $e_{\rm CJ}$ in our study is likely produced by scattering among giant planets after the gas disk has dissipated \citep{2025ApJ...980L..31W}, we introduce a CJ into the numerical integrations only after turning off the dissipative forces responsible for planet migration in a gaseous disk. In reality, however, a CJ should have formed within a protoplanetary disk in order to accrete a substantial gaseous envelope. 
While the presence of a CJ could affect the resonance libration of the inner SEs (see Appendix \ref{sec:capture}), the configurations of the pairs that do capture into resonance are established by orbital decay and eccentricity damping, prior to the onset of secular forcing by the CJ. Our approach, therefore, does not differ qualitatively for the systems of interest.

The effects of general relativity, the stellar quadropole potential, tidal dissipation, and planet-planet collisions are not considered; stars and planets are treated as point particles in this study.  Consequently, predicting the ultimate fate of SE1, SE2, and the cold Jupiter in the subsequent evolutions, such as their final orbital locations and scattering/merging events, is outside the scope of our present statistical study. Nevertheless,
our goal is to investigate the chance of a cold Jupiter exciting the resonance libration and TTV phase of the inner planetary pair or, more violently, the possibility of disrupting the resonance. Resonance disruption in this study is defined as $|\Delta| > 10$\%, i.e., a significant deviation from the observed value (and from our initial values) of  $\lesssim$ a few percent.

\begin{figure*}[ht!]
\epsscale{1.16}
\plotone{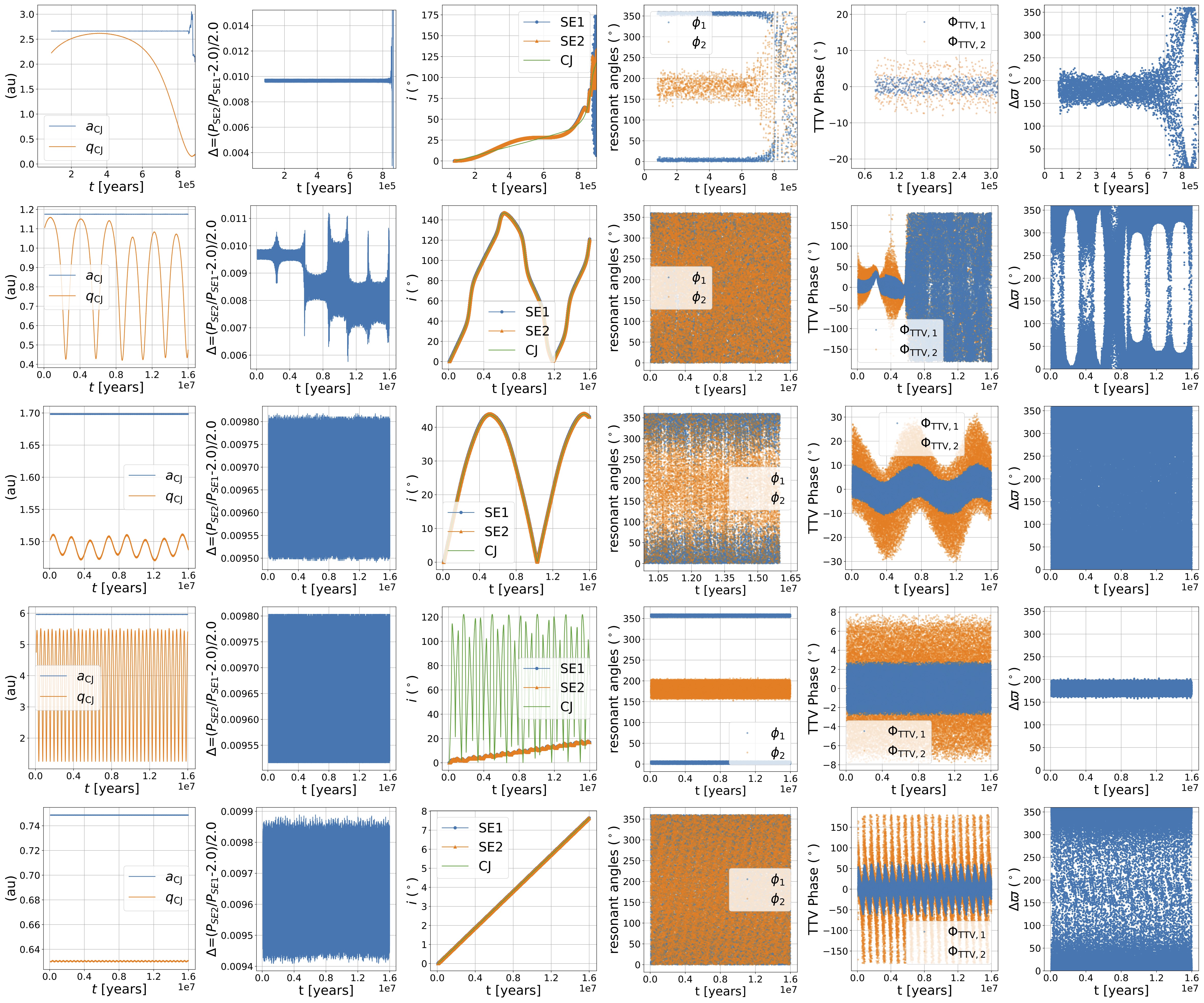}
\caption{An evolutionary gallery of five examples of planetary systems over a timespan of 16 Myr: extreme EKL (top row), strong EKL (second row), mild EKL (third row), weak EKL (fourth row), and decoupled EKL (bottom row). Each row illustrates the evolution of the semimajor axis and periapsis distance of the cold Jupiter ($a_{\rm CJ}$, $q_{\rm CJ}$), the fractional deviation from exact mean-motion commensurability $\Delta$, the orbital inclinations of the planets ($i$), the resonant angles ($\phi_1$, $\phi_2$), the TTV phases of SE1 and SE2 ($\Phi_{\rm TTV,1}$, $\Phi_{\rm TTV,2}$), and the difference in the longitudes of periapsis between SE1 and SE2 ($\Delta \varpi$). The extreme EKL case is shown only for the first 0.9 Myr, up to the disruption of the resonant planetary pair. For the mild EKL case, the resonant angles are plotted only for $t>10^7$ yr in order to zoom in and clearly distinguish the evolution of the two angles.
\label{fig:r013_EKLexample}}
\end{figure*}

\begin{deluxetable}{lccccccc}
\tablecaption{Initial condition for the example runs illustrated in Figure \ref{fig:r013_EKLexample}.The last column also shows the minimum $q_{\rm CJ}$ during the evolutions.\label{table_fig1}}
\tablehead{
\colhead{EKL type} & \colhead{$a_{\rm CJ}$ (au)} &  \colhead{$e_{\rm CJ}$}  & \colhead{$m_{\rm CJ}$ ($M_{\rm jup}$)} &  \colhead{$a_{\rm star2}$ (au)} & \colhead{$e_{\rm star2}$} &
\colhead{$i_{\rm star2}$ ($^\circ$)} &
\colhead{min($q_{\rm CJ}$) (au)}
}
\startdata
extreme & 2.66  & 0.16 &  1.42 & 176.4 & 0.23 & 75 & 0.15 \\
strong & 1.17 & 0.58 & 7.48 & 292.4 & 0.76 & 67.5 & 0.43\\
mild & 1.7  & 0.11  & 4.77 & 280.4 & 0.53 & 158 & 1.46 \\
weak & 5.96 & 0.2 & 0.4 & 206.2 & 0.27 & 119 & 1.25 \\
decoupled & 0.75 & 0.158 & 0.89 & 490 & 0.02 & 67.9 & 0.63 \\
\enddata
\end{deluxetable}

\section{Example Runs}
\label{sec:example}

Figure \ref{fig:r013_EKLexample} illustrates the orbital evolution of five example runs in the r0.13 model for the 2:1 resonant SE pair, presenting five scenarios as described in the figure caption: extreme EKL, strong EKL, mild EKL, weak EKL, and decoupled EKL perturbations. The initial conditions for these five examples are listed in Table \ref{table_fig1}. The last column of Table \ref{table_fig1} also shows the minimum of the periapsis distance $q_{\rm CJ}$ during the evolutions. As we shall see, the SE pairs are nearly coplanar during evolution. Hence, we adopt the analytic theory of \citet{2012ApJ...761..122L} to compute the TTV phases (hereafter the L12 model; see Appendix \ref{sec:appendix}).

In the first example shown in the top row, the cold Jupiter reaches $q_{\rm CJ}\approx 0.15$ au during the first EKL cycle. As indicated by the corresponding plots, this extremely close periastron passage triggers strong non-secular gravitational interactions, which disrupt the resonant planetary pair and lead to substantial variations in $a_{CJ}$, the separation parameter $\Delta$, the planetary orbital inclinations ($i$), the resonant angles ($\phi_{1}$ and $\phi_{2}$), and the mutual longitude of periapsis of the planetary pair ($\Delta \varpi$). The TTV phases ($\Phi_{\rm TTV}$) are displayed only up to a time well before this close encounter, beyond which double transits are unlikely due to the large mutual inclination acquired by the SE pair. Note that the initial inclination of the companion star is already relatively high, $\approx 75^\circ$.

The second row of Figure \ref{fig:r013_EKLexample} illustrates a case of strong EKL perturbations. 
The stellar-induced EKL oscillations drive the cold Jupiter to a minimum periastron distance of $q_{\rm CJ}\approx 0.43$ au. During this evolution, $\Delta$ varies within the interval $\approx 0.006$–0.011, which is wider than its initial range $\approx 0.0095$-0.0098. Although the resonance between the SE pair is not disrupted, these close encounters substantially amplify the TTV phases ($|\Phi_{\rm TTV,1}|=|\Phi_{\rm TTV,2}|\approx 180^\circ$) and induce circulation of the resonant angles. The EKL-driven evolution also significantly increases the planets’ orbital inclinations relative to the initial coplanar configuration ($i=0$), while the cold Jupiter remains nearly coplanar with the SE pair. This behavior arises because the precession of the SE pair induced by the cold Jupiter proceeds on a much shorter timescale than the precession of the cold Jupiter induced by the stellar companion, resulting in a stronger dynamical coupling between the SE pair and the cold Jupiter than between the stellar companion and the cold Jupiter \citep{2017AJ....153...42L}. This hierarchy in dynamical coupling strengths is also reflected in the relatively slow EKL oscillation of the cold Jupiter in this case. In addition, the resonant SE pair can become temporarily apsidally aligned during each close approach of the cold Jupiter, owing to strong secular perturbations from the eccentric third body that intermittently break the resonance lock \citep{2023MNRAS.522.1914C}. In this example, the SE pair is rendered dynamically hot as a consequence of the strong EKL-driven interactions.

The third row of Figure \ref{fig:r013_EKLexample} shows an example of mild EKL perturbations.
In this case, the stellar EKL mechanism drives the cold Jupiter to a periapsis distance of $q_{\rm CJ}\approx 1.46$ au. Compared to the strong EKL case, $\Delta$ exhibits only modest fluctuations around its initial value throughout the simulation. All planets remain nearly coplanar, consistent with the slow EKL oscillations. Because the EKL perturbation arises from an external companion relative to the resonant pair, the resonant angle $\phi_1$ remains in a state of near-libration, whereas $\phi_2$ transitions to circulation. The TTV phases $\Phi_{\rm TTV,1}$ and $\Phi_{\rm TTV,2}$ can attain amplitudes of approximately $10^\circ$ and $30^\circ$, respectively. The mild EKL perturbations also disrupt the initial anti-alignment of the planetary periapses but do not drive the system toward the extreme apsidal alignment observed in the strong EKL case.

The planetary system in the fourth row presents an example of weak EKL perturbations. 
The system 
exhibits rapid EKL-driven oscillations in the eccentricity (thus $q_{\rm CJ}$) and in the inclination $i$ of the cold Jupiter with respect to the initial orbital plane. 
Nevertheless, $\Delta$ retains its initial range of values. SE1 and SE2 remain nearly coplanar, whereas the orbital plane of the cold Jupiter becomes significantly inclined relative to the orbits of the SE pair owing to the stronger dynamical coupling between the cold Jupiter and the stellar companion, consistent with the observed rapid EKL oscillations.\footnote{Quantitatively, the ratio of the precession rate of the cold Jupiter induced by the stellar companion to that of the SE pair induced by the cold Jupiter, $\sim (m_{\rm star2}/m_{\rm CJ})a_{\rm CJ}^{9/2}/(a_{\rm star2}^3 a_{\rm SE}^{3/2})$ \citep{2017AJ....153...42L}, is $\sim 20$ in this example, whereas this ratio is $\sim 2\times 10^{-4}$ in the strong EKL example.} Despite frequent EKL oscillations, the minimum periapsis distance of the cold Jupiter, $\min(q_{\rm CJ})$, remains beyond 1 au. Consequently, the cold Jupiter fails to significantly excite the TTV phases (with $\max|\Phi_{\rm TTV,1}|\approx 2.7^\circ$ and $\max|\Phi_{\rm TTV,2}|\approx 8^\circ$) or the libration amplitudes of the resonant angles of the SE pair (with $\max(\Delta \phi_1) \approx 14^\circ$ and $\max(\Delta \phi_2) \approx 40^\circ$). The EKL perturbation is therefore weak in this case: the periapses of the resonant pair remain nearly anti-aligned ($\Delta \varpi \approx 180^\circ \pm 20^\circ$), and the pair stays trapped in resonance for 16 Myr. The SE pair is dynamically colder than in the preceding mild EKL example, primarily because the cold Jupiter mass is substantially smaller in the present case ($m_{\rm CJ}\approx 0.4\,M_{\rm Jup}$, compared to $m_{\rm CJ}\approx 4.77\,M_{\rm Jup}$ in the previous example), leading to correspondingly weaker EKL perturbations.

Finally, the bottom row of Figure \ref{fig:r013_EKLexample} illustrates an example decoupled from EKL interactions because the initial EKL cycle period $t_{\rm EKL}$ is larger than the precession time of the cold Jupiter induced by the J$_2$ gravitational potential of the resonant SE pair $t_{\rm SE,J_2}$ (see Appendix \ref{sec:timescale}). 
Therefore,
$q_{\rm CJ}$ and $i$ do not undergo significant EKL-driven oscillations, and thus the planetary system is dynamically decoupled from the stellar EKL mechanism. All planets remain nearly coplanar, and their orbital planes slowly drift away from the initial plane by only $\sim 8^\circ$ over 16 Myr due to the inclined stellar companion. The resonant offset $\Delta$ fluctuates within a range barely larger than its initial values. Nevertheless, thanks to the small $q_{\rm CJ}$, the cold Jupiter is capable of inducing circulation of the resonant angles and can intermittently excite $\Phi_{\rm TTV,1}$ and $\Phi_{\rm TTV,2}$ up to $\sim 60^\circ$ and $\sim 180^\circ$, respectively. The resonant pair becomes mildly apsidally aligned, i.e., $\Delta \varpi$, librates around alignment with comparatively large amplitudes, driven by the modest secular perturbations from the cold Jupiter.

\section{Statistical Results}
We conduct N-body simulations of ensembles of planetary systems subject to the EKL mechanism from a stellar companion. To isolate and characterize the impact of the EKL mechanism, we first analyze results from our model, r0.13, in the absence of a stellar companion (hereafter r0.13 w/o EKL). 
We then examine the same model in the presence of a stellar companion to contrast the results and thereby distinguish the EKL effect. 
We also supplement a study of 
the ``circ" model from \citet{2025ApJ...980L..31W}, in which the initial eccentricity of a cold Jupiter is small ($e_{\rm CJ} = 0.01$), for further comparison. 
Each ensemble comprises 200 runs (i.e., 200 planetary systems), initialized from the distributions of orbital and physical parameters outlined in Section \ref{sec:setup}. The systems are numerically integrated for 16 Myr to enable a statistical investigation of TTV-phase and libration-amplitude excitation, as well as resonance disruption.

As illustrated in Section \ref{sec:example}, all orbital parameters, including TTV phases and the libration amplitude of resonant angles, exhibit temporal variations. To infer their statistical properties and characterize dynamical activity, we shall refer to their extreme values during evolution, such as the maximum absolute value of TTV phases (max($|\Phi_{\rm TTV}|$)), the maximum libration amplitude of renonant angles (max($\Delta \phi$)), and the minimum value of the periapsis of the cold Jupiter  (min($q_{\rm CJ}$)). Most of the time, $|\Phi_{\rm TTV}|$ and $\Delta \phi$ typically fluctuate within these extreme values, or circulate when max($|\Phi_{\rm TTV}|)=180^\circ$ and max($\Delta \phi ) =360^\circ$.

\subsection{The r0.13 model in the absence of a stellar companion}
\label{sec:r0.13_noEKL}

\begin{figure}[t]
    \centering
    \begin{minipage}{0.23\textwidth}
        \centering
        \includegraphics[width=\linewidth]{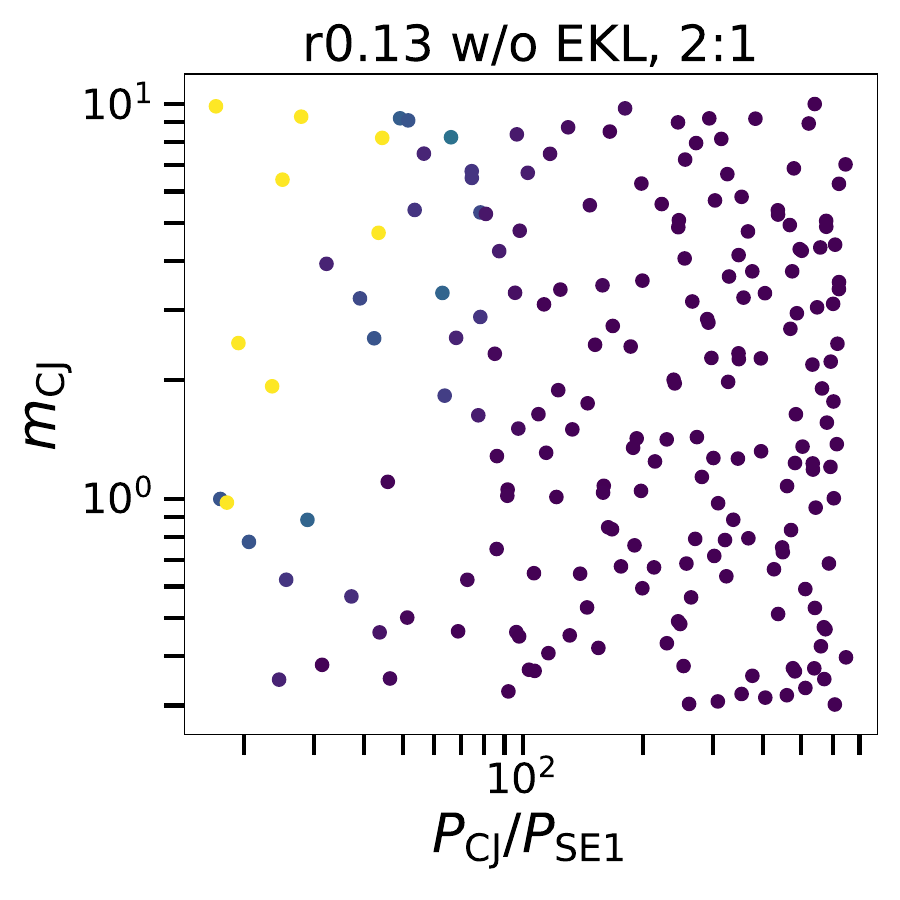}
    \end{minipage}
    \hspace{-0.017\textwidth}
    \begin{minipage}{0.23\textwidth}
        \centering
        \includegraphics[width=\linewidth]{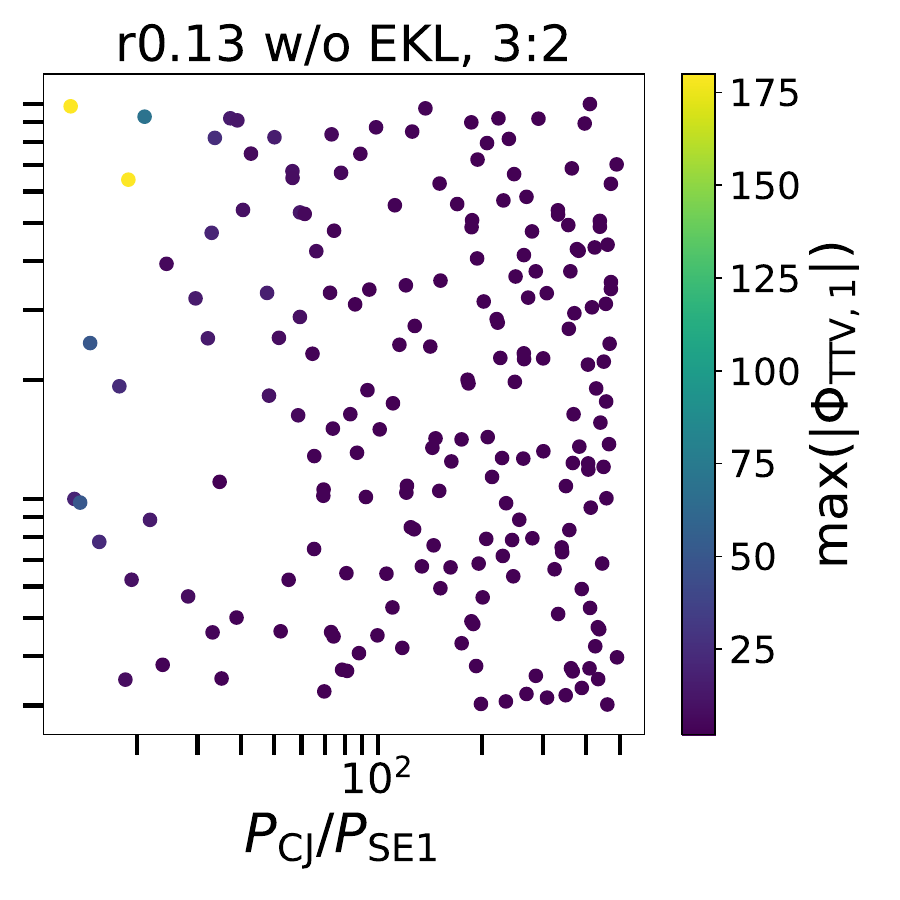}
    \end{minipage}

    \vspace{0 em} 

    \begin{minipage}{0.23\textwidth}
        \centering
        \includegraphics[width=\linewidth]{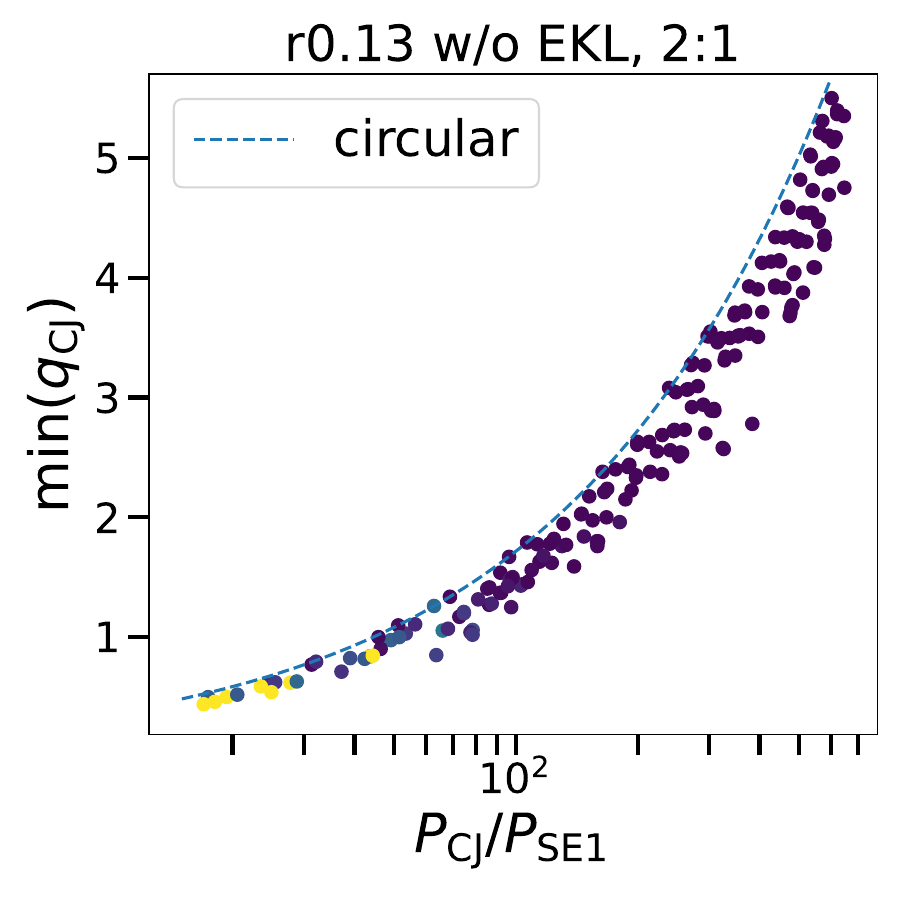}
    \end{minipage}
    \hspace{-0.017\textwidth}
    \begin{minipage}{0.23\textwidth}
        \centering
        \includegraphics[width=\linewidth]{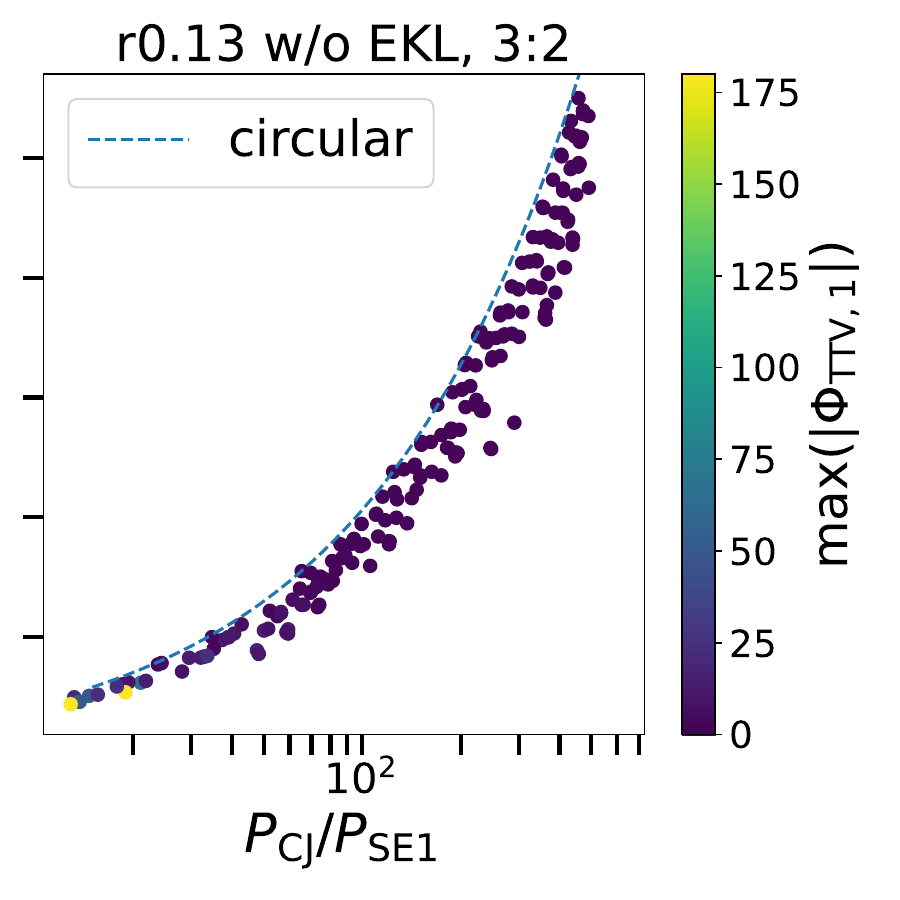}
    \end{minipage}

    \vspace{0 em}

    \begin{minipage}{0.23\textwidth}
        \centering
        \includegraphics[width=\linewidth]{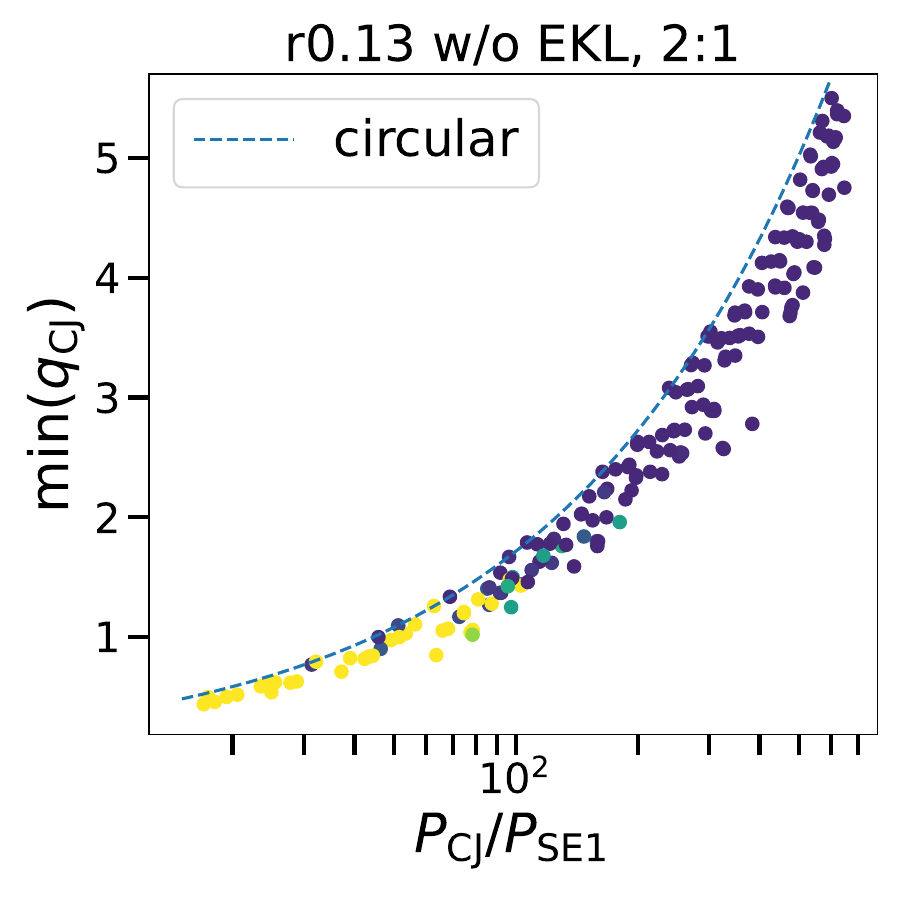}
    \end{minipage}
    \hspace{-0.017\textwidth}
    \begin{minipage}{0.23\textwidth}
        \centering
        \includegraphics[width=\linewidth]{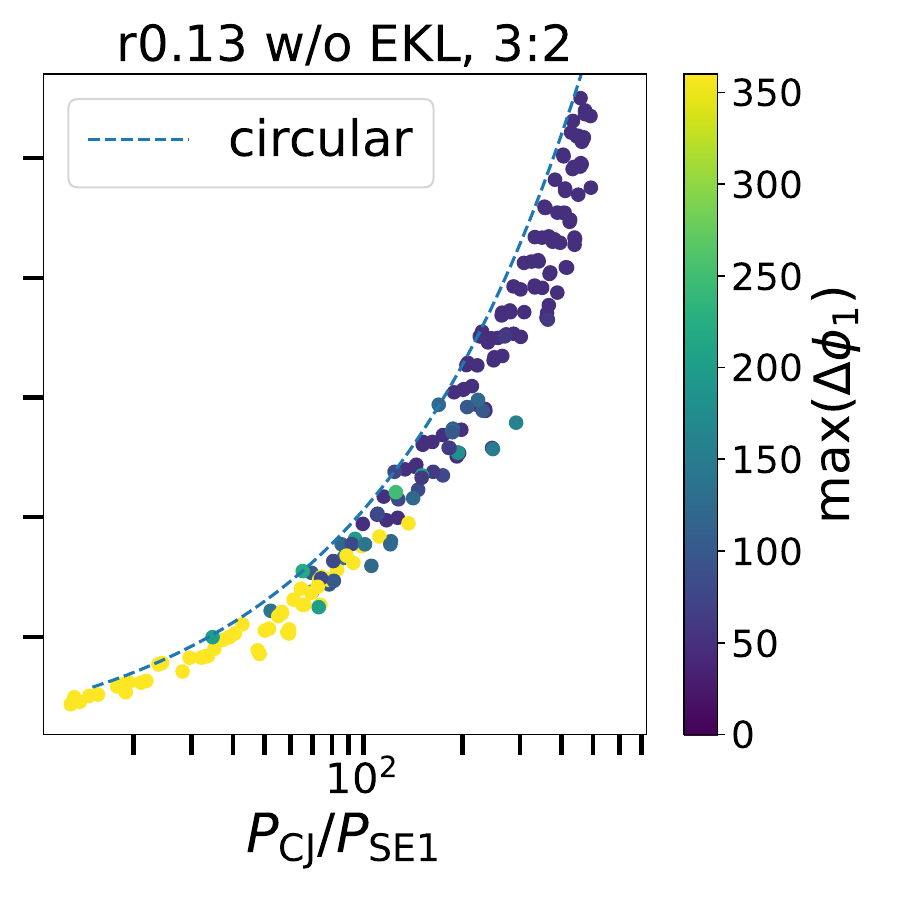}
    \end{minipage}

    \caption{Comparison of $\max\left(|\Phi_{\rm TTV,1}|\right)$ and $\max(\Delta \phi_1)$, color-coded, for 2:1 (left panels) and 3:2 (right panels) resonant pairs in the r0.13 model in the absence of a stellar companion. The distribution of $\max\left(|\Phi_{\rm TTV,1}|\right)$ is presented in relation to the cold Jupiter mass $m_{\rm CJ}$ and the period ratio $P_{\rm CJ}/P_{\rm SE1}$ (top panels), and as a function of $\min(q_{\rm CJ})$ and the same period ratio (middle panels). The distribution of $\max(\Delta \phi_1)$ as a function of $\min(q_{\rm CJ})$ and $P_{\rm CJ}/P_{\rm SE1}$ is presented in the bottom panels. For reference, the dashed curve, denoted by ``circular'', corresponding to $e_{\rm CJ}=0$, is overplotted in the middle and bottom panels to illustrate how the simulated distribution of $\min(q_{\rm CJ})$ deviates from circular orbits.}
    \label{fig:r013_noEKL}
\end{figure}

Figure \ref{fig:r013_noEKL} presents the distributions of the maximum TTV phase of SE1, max($|\Phi_{\rm TTV,1}|$), and the maximum libration amplitude of the resonant angle of SE1, max($\Delta \phi_1$), in the r0.13 model in the absence of a stellar companion. Each data point corresponds to one planetary system and is color-coded to indicate its values of max($|\Phi_{\rm TTV,1}|$) and max($\Delta \phi_1$). The top panels display the distribution in the parameter space of the cold Jupiter mass $m_{\rm CJ}$ vs. the period ratio $P_{\rm CJ}/P_{\rm SE1}$. These plots are intended to compare with the results from \citet{2023MNRAS.522.1914C} for TTV phase excitation by a coplanar cold Jupiter. 
The distribution plots show that max($|\Phi_{\rm TTV}|$) increases from a low-mass cold Jupiter at a large orbit to a high-mass cold Jupiter at a small orbit in both resonant-pair cases. This trend is consistent with that shown in Figure 12 of \citet{2023MNRAS.522.1914C}. Owing to the initial eccentricity following the Rayleigh distribution with a small mean of 0.13, this ensemble of cold Jupiters is expected to remain at low eccentricities without a stellar perturbation \citep{2025ApJ...980L..31W}. Indeed, the small eccentricities of cold Jupiters are illustrated in the middle panels of Figure \ref{fig:r013_noEKL}, where most data points lie close to the curve for circular orbits (denoted by ``circular") with a small spread in periapsis $q_{\rm CJ}$. 
Consequently, no cold Jupiters can excite large TTV phases in the lower-right part of the parameter space of $m_{\rm CJ}$ vs. $P_{\rm CJ}/P_{\rm SE1}$; their small eccentricity cannot lead to smaller periapsiss to secularly excite significant TTV phases of close-in resonant pairs if they are less massive and farther away. Unlike $|\Phi_{\rm TTV,1}|$, $\Delta \phi_1$ of SE1 with close-in cold Jupiters can be excited to larger values $\sim 360^\circ$, i.e., becoming circulation, as shown in the bottom panels of Figure \ref{fig:r013_noEKL}. 

As illustrated by Figure \ref{fig:r013_noEKL}, the closer-in cold Jupiters, with the period ratio $P_{\rm CJ}/P_{\rm SE1} \lesssim 100$ for the 2:1 pair ($\lesssim 30$ for the 3:2 pair), can excite moderate and large TTV phases, depending on $m_{\rm CJ}$. Since the initial Rayleigh distribution of cold Jupiter's eccentricity may arise from planet-planet scattering after the dispersal of the disk \citep{2025ApJ...980L..31W}, these dynamically hotter pairs associated with closer-in cold Jupiters may resemble a larger TTV phase and resonant-angle libration/circulation by a cold Jupiter diving into a smaller orbit through planet-planet scattering.

\begin{figure}[ht]
    \centering
    \begin{minipage}{0.23\textwidth}
        \centering
        \includegraphics[width=\linewidth]{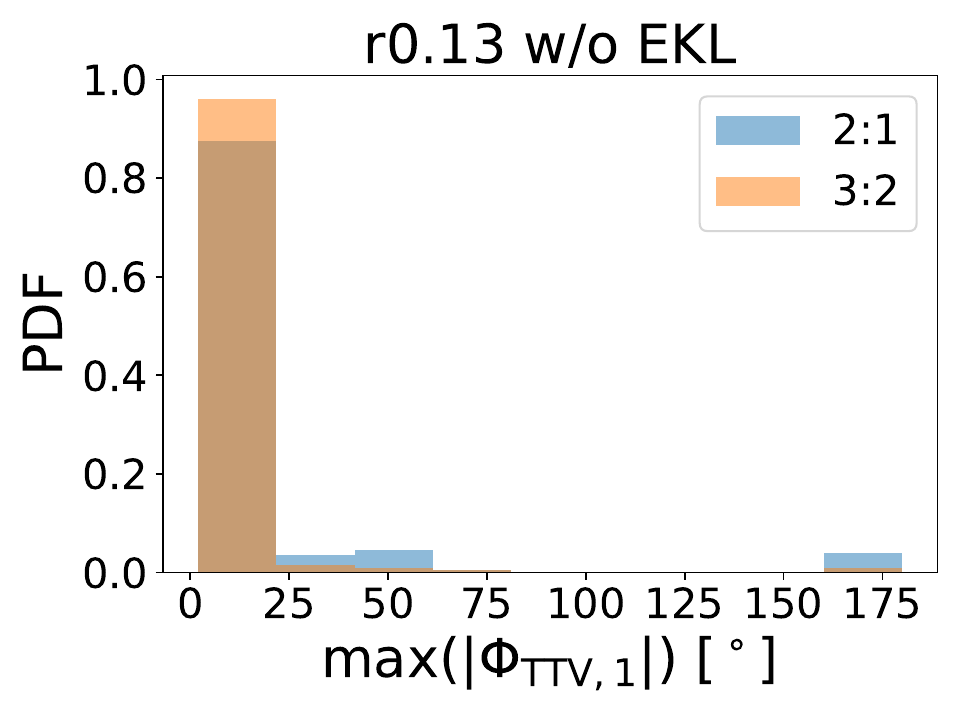}
    \end{minipage}
    \hfill
    \begin{minipage}{0.23\textwidth}
        \centering
        \includegraphics[width=\linewidth]{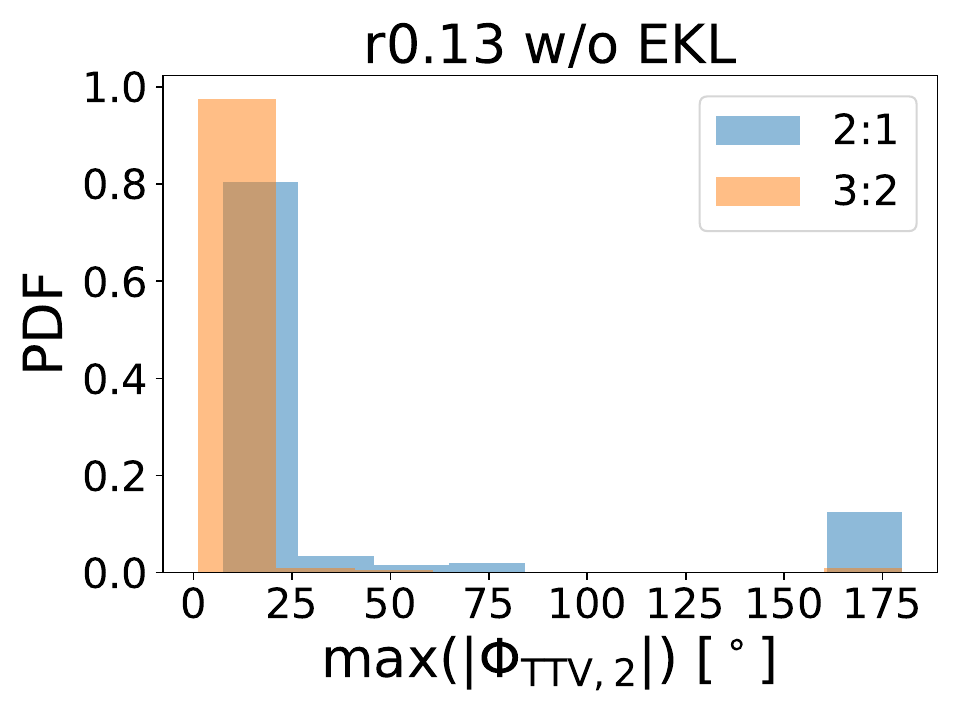}
    \end{minipage}

    \vspace{0 cm} 

    \begin{minipage}{0.23\textwidth}
        \centering
        \includegraphics[width=\linewidth]{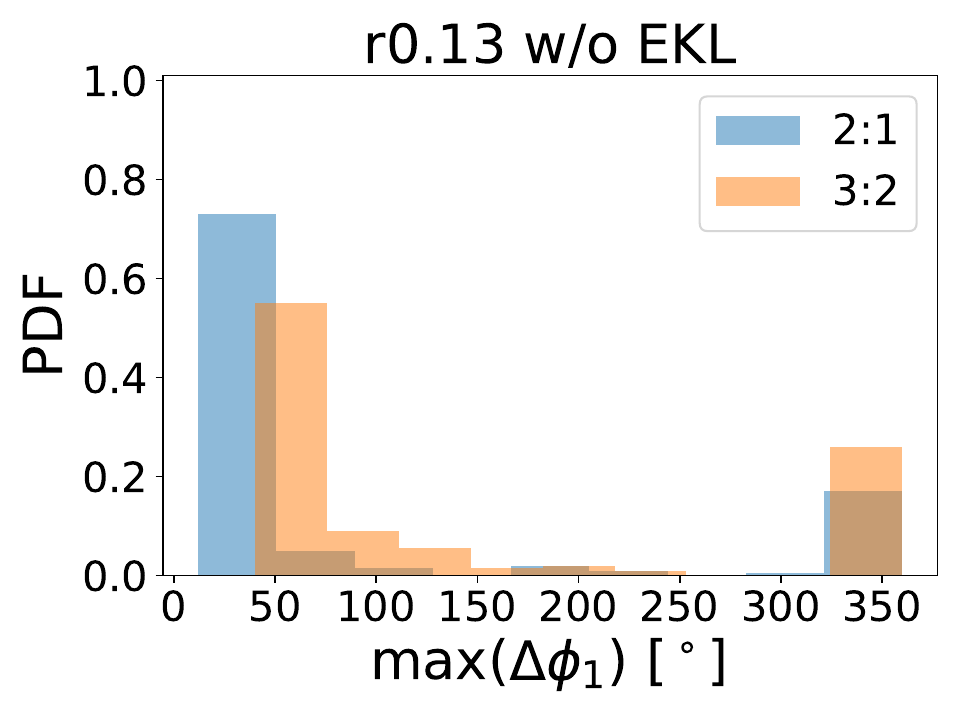}
    \end{minipage}
    \hfill
    \begin{minipage}{0.23\textwidth}
        \centering
        \includegraphics[width=\linewidth]{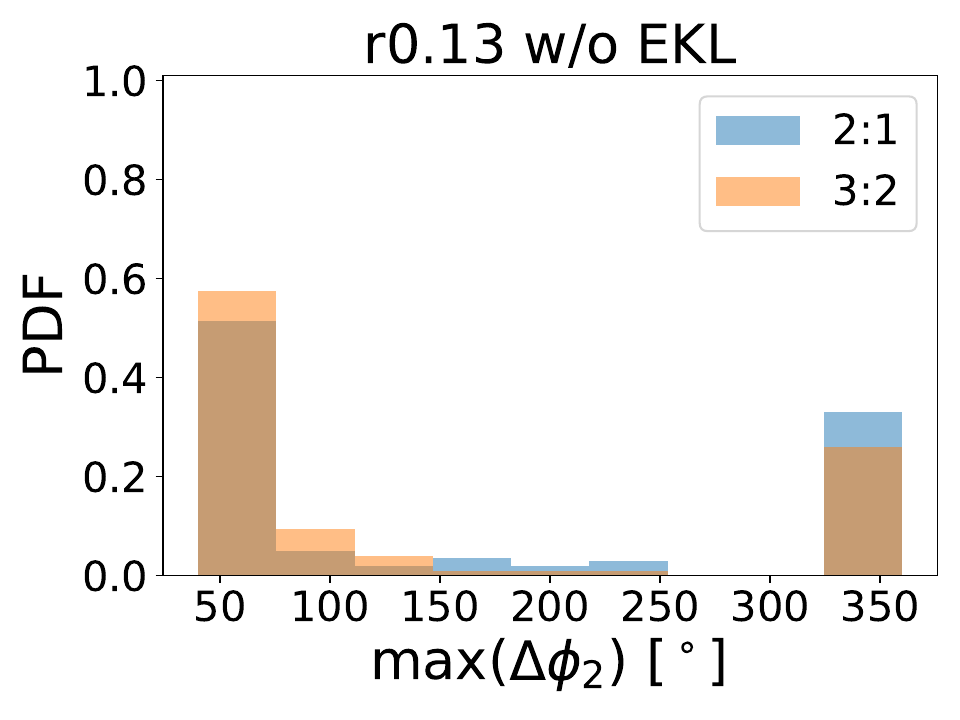}
    \end{minipage}
    
    \caption{Histogram of the probability distribution function (PDF) of max($|\Phi_{\rm TTV,1}|$), max($|\Phi_{\rm TTV,2}|$), max($\Delta \phi_1$), and max($\Delta \phi_2$) in the r0.13 model in the absence of a stellar companion. The cases for the 2:1 and 3:2 resonances are shown for comparison. There is a deficit for max($|\Phi_{\rm TTV,1}|$) and max($|\Phi_{\rm TTV,2}|$) $\lesssim 25^\circ$ and an excess for them $> 25^\circ$ in the 2:1 resonance compared to the 3:2 one. The PDFs of max($\Delta \phi_1$) and max($\Delta \phi_2$) are similar for the 3:2 resonance, while they differ noticeably for the 2:1 resonance.}
    \label{fig:histr013_noEKL}
\end{figure}

As has been explained by \citet{2023MNRAS.522.1914C}, the 2:1 pair is less resistant to TTV phase excitations by a third-body secular perturbation than the 3:2 pair, due to the indirect potential that weakens resonant interactions in the 2:1 resonance. This feature is observed in Figure \ref{fig:r013_noEKL}, where more planets have larger max($|\Phi_{\rm TTV,1}|$) in 2:1 resonance than in 3:2 resonance. Specifically, the histograms of the probability distribution function (PDF) in the top panels of 
Figure \ref{fig:histr013_noEKL} shows that more 2:1 pairs have larger TTV phases than 3:2 pairs, consistent with the expectation from the relative resonant strength. The similar feature is also seen for max($\Delta \phi_2$) in the lower right panel of Figure \ref{fig:histr013_noEKL}, reflecting the same dynamics.  max($\Delta \phi_1$) is less excited by a cold Jupiter in the 2:1 resonance, as shown in the lower left panel, due to the weaker resonant coupling in the 2:1 resonance. In contrast, the PDFs of max($\Delta \phi_1$) and max($\Delta \phi_2$) in the 3:2 resonance are almost identical.

Overall, while the resonant angle of $\sim 20$-30\% of resonant pairs end up with circulating due to the secular perturbation of a cold Jupiter, it appears difficult for both 2:1 and 3:2 pairs to gain the TTV-phase excitation $\sim 180^\circ$ for 16 Myr: $<$ 12.5\% in our ensemble of 200 planetary systems. This is because circulating resonant angles occur as long as the free eccentricity is not smaller than the forced eccentricity, whereas circulating TTV phases require the free eccentricity to be larger than the forced eccentricity (see Appendix \ref{sec:appendix}).
 
\subsection{The r0.13 model in the presence of a stellar companion}
\label{sec:r0.13}
\begin{figure}[ht!]
\epsscale{1.2}
\plotone{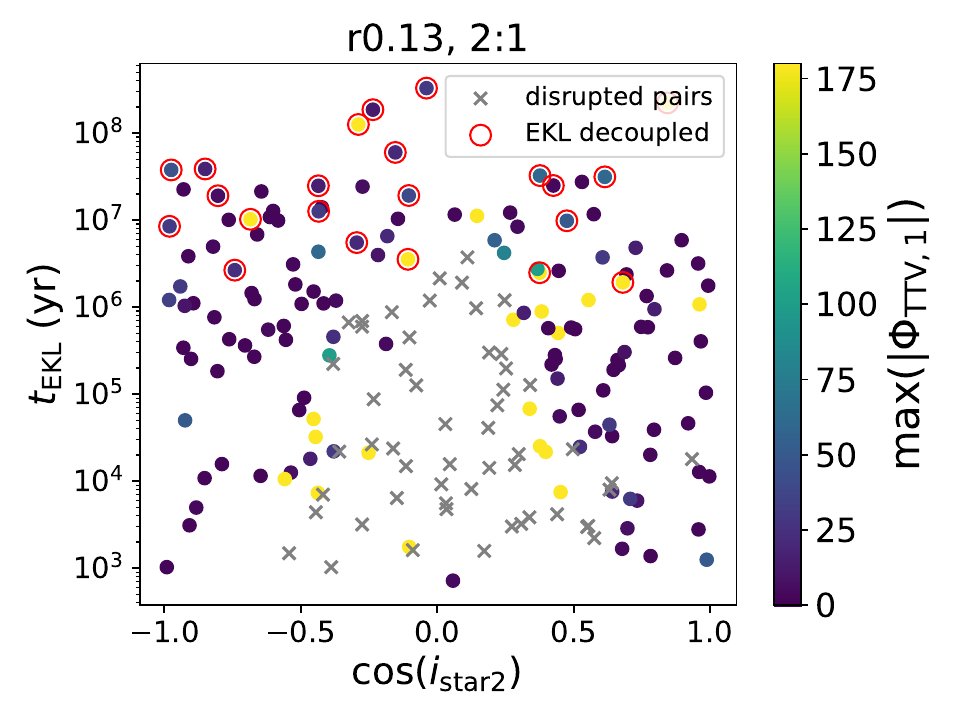}
\caption{Distribution of the maximum of TTV phase of SE1, max($|\Phi_{\rm TTV,1}|$) (color-coded), and resonant disruption (denoted by a grey cross symbol) in terms of the initial EKL timescale $t_{EKL}$ and orbital inclination of stellar companion $\cos i_{\rm star2}$. The data points surrounded by a red circle indicate the systems decoupled dynamically from the stellar companion because $t_{\rm EKL} > t_{\rm SE,J_2}$. The distribution is illustrated for the 2:1 resonance in the r0.13 model. A similar distribution appears for the 3:2 resonance (not shown).
\label{fig:figr013_EKL_disr}}
\end{figure}

In the presence of a stellar companion, the eccentricity and inclination oscillations driven by the EKL mechanism occur provided $t_{\rm EKL}<t_{\rm SE,J_2}$, or equivalently, provided the hierarchy parameter $\epsilon \equiv a_{\rm CJ}e_{\rm star2}/a_{\rm star2}/(1-e_{\rm star2}^2)$ is large enough to ignore the SE gravitational potential. One striking feature of introducing stellar EKL is the disruption of resonances, which never occurs in the model without a stellar companion, as presented in the previous section.  Figure \ref{fig:figr013_EKL_disr} depicts the parameter space allowing for resonance disruption or large TTV phase amplitudes to occur, in terms of initial $t_{\rm EKL}$ and $i_{\rm star2}$, in the r0.13 model for the 2:1 resonant pairs. The resonance disruption, denoted by grey cross symbols, predominantly populates the parameter space where the initial inclination angles are large, i.e., $|\cos(i_{\rm star2})| \lesssim 0.57$. This result is consistent with a binormal distribution of mutual inclinations of cold Jupiters, with peaks near $55^\circ$ and $125^\circ$ and a paucity in between, in the r0.13 model of \citet{2025ApJ...980L..31W}. It is because the initial large mutual inclinations tend to drive extreme EKL oscillations, which then disrupt a resonant pair due to the close approach of a cold Jupiter on a highly eccentric orbit. The ensemble of the planetary system for max($|\Phi_{\rm TTV,2}|$) and for the 3:2 resonance also exhibits a distribution very similar to that presented in Figure \ref{fig:figr013_EKL_disr}, so they are not shown here.

The planetary systems with non-disruptive pairs are populated distinctly in the rest of the parameter space in Figure \ref{fig:figr013_EKL_disr}.
Generally speaking, including EKL allows a larger parameter of space with $t_{\rm EKL}<t_{\rm SE,J_2}$ to drive moderate and large TTV phase amplitudes (max($|\Phi_{\rm TTV}|) \gtrsim 25^\circ$).
Although the systems coupled with a stellar companion on a modestly inclined orbit tend to avoid resonance disruption, the EKL can excite moderate and large TTV phases of SE pairs. In comparison, the systems with a closer-in cold Jupiter, despite being EKL decoupled from a stellar companion, can still excite significant TTV phases, as in the cases with closer-in cold Jupiters in the absence of a stellar companion shown in Section \ref{sec:r0.13_noEKL}.

\begin{figure}[ht]
    \centering
    \begin{minipage}{0.23\textwidth}
        \centering
        \includegraphics[width=\linewidth]{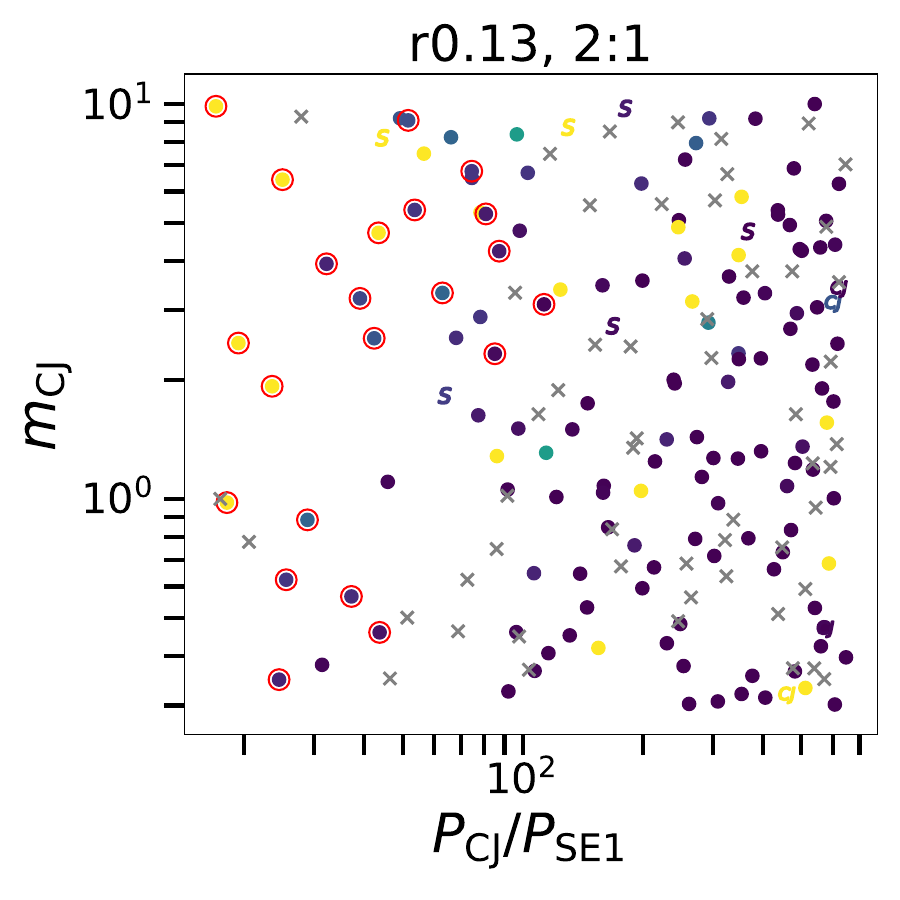}
    \end{minipage}
    \hspace{-0.017\textwidth}
    \begin{minipage}{0.23\textwidth}
        \centering
        \includegraphics[width=\linewidth]{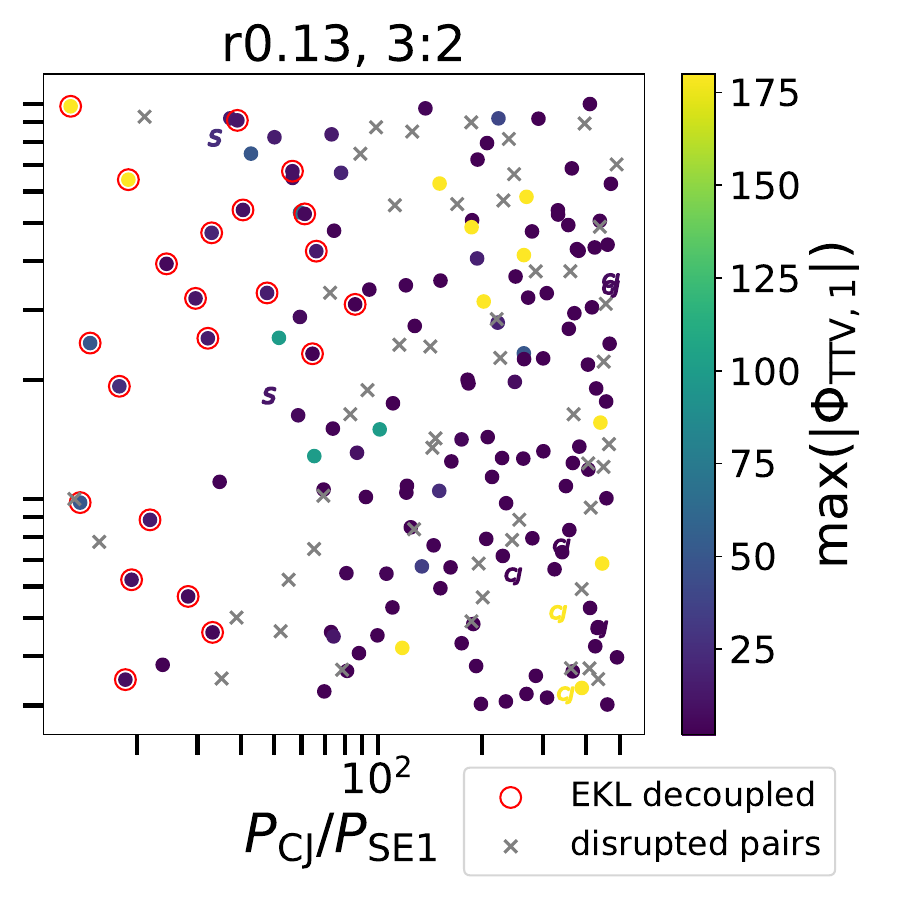}
    \end{minipage}

    \vspace{0 em} 

    \begin{minipage}{0.23\textwidth}
        \centering
        \includegraphics[width=\linewidth]{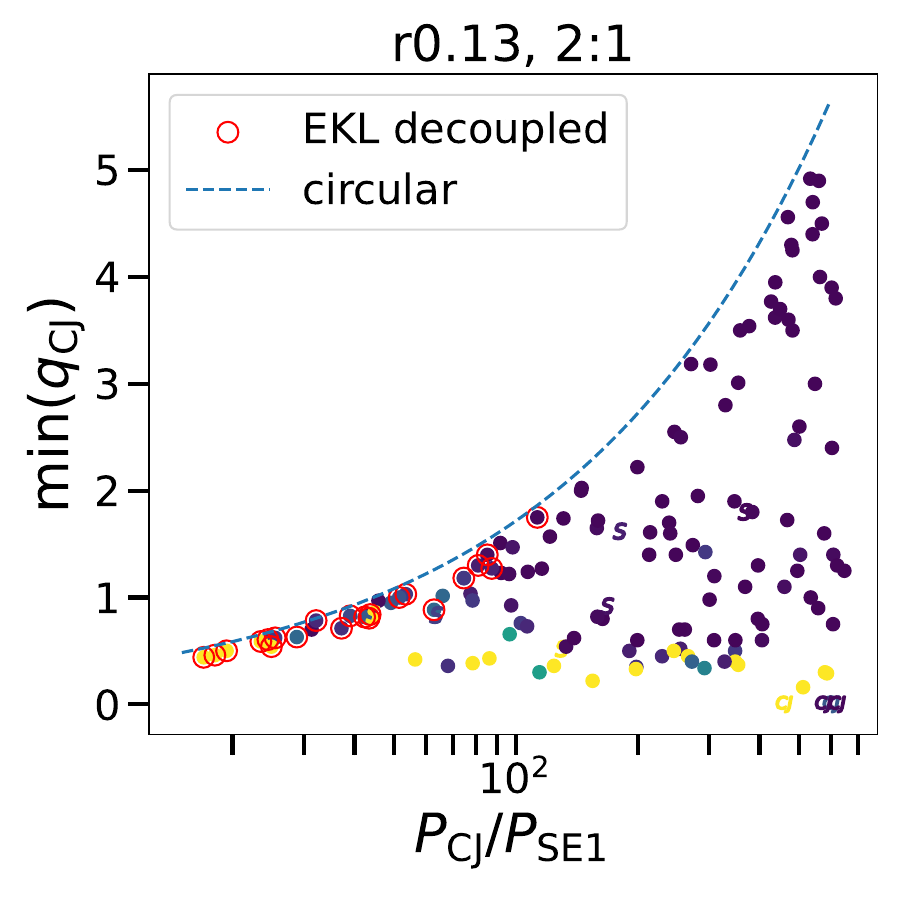}
    \end{minipage}
    \hspace{-0.017\textwidth}
    \begin{minipage}{0.23\textwidth}
        \centering
        \includegraphics[width=\linewidth]{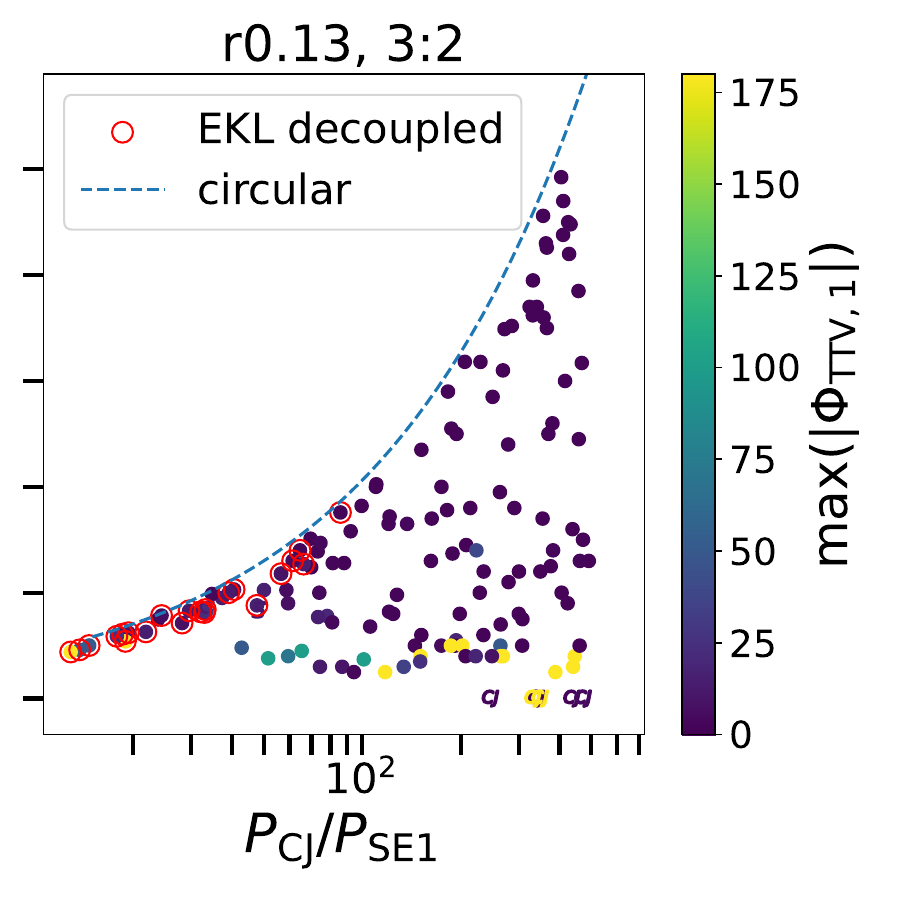}
    \end{minipage}

    \vspace{0 em}

    \begin{minipage}{0.23\textwidth}
        \centering
        \includegraphics[width=\linewidth]{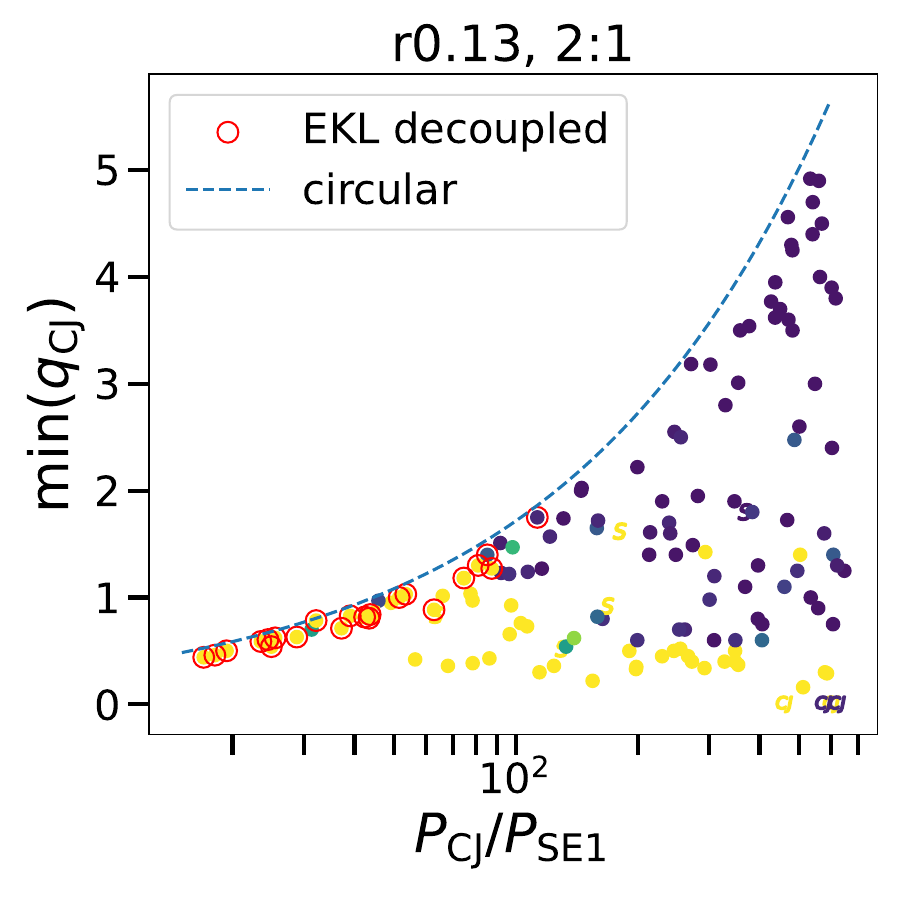}
    \end{minipage}
    \hspace{-0.017\textwidth}
    \begin{minipage}{0.23\textwidth}
        \centering
        \includegraphics[width=\linewidth]{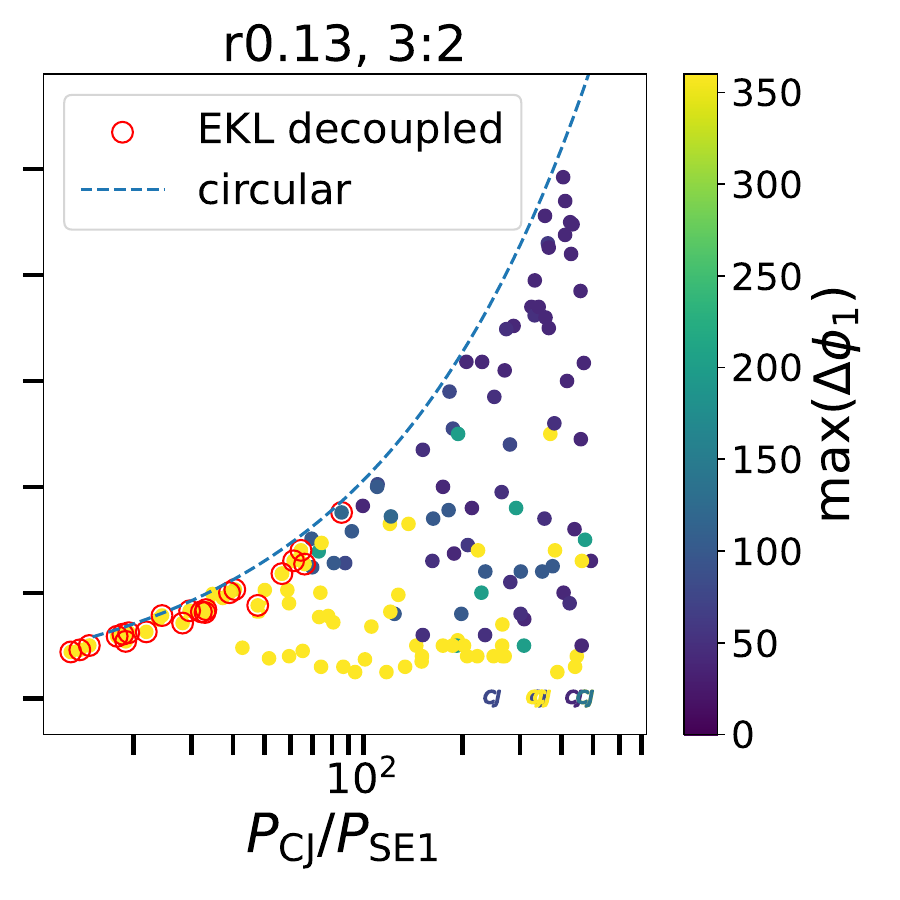}
    \end{minipage}

    \caption{Same as Figure \ref{fig:r013_noEKL} for the ``r0.13 w/o EKL" model, but with a stellar companion . Data points labeled $CJ$ and $S$ (instead of solid circles) mark systems where a cold Jupiter is ejected or the companion star becomes quasi-unstable, respectively.\footnote{Please read the electronic version to read them clearly.} Only disrupted resonant pairs are shown in the top panels. Relative to Figure \ref{fig:r013_noEKL}, the distributions of large max($|\Phi_{\rm TTV,1}|$) and max($\Delta \phi_1$) values are much more dispersed in these parameter spaces due to EKL oscillations.}
    \label{fig:r013_EKL}
\end{figure}

Figure \ref{fig:r013_EKL} further illustrates EKL excitation of orbital eccentricities of cold Jupiters and the TTV phase of SE1 in planetary systems without resonant disruption. Comparing the middle and bottom panels of Figure \ref{fig:r013_noEKL} with those of Figure \ref{fig:r013_EKL}, we clearly see that stellar EKL excites cold-Jupiter eccentricities from the initial Rayleigh distribution, populating a broad range of min($q_{\rm CJ}$) below the $e_{\rm CJ}=0$ curve denoted by ``circular". More specifically, the minimum periapsis, min($q_{\rm CJ}$), of distant cold Jupiters -- $P_{\rm CJ}/P_{\rm SE1} > 50$ -- can also become small, driving TTV phases due to large eccentricities excited by EKL. According to Equations (\ref{eq:t_EKL}) and (\ref{eq:t_J2}), $P_{\rm CJ}/P_{\rm SE1} \sim 50$ roughly corresponds to $t_{\rm EKL} \sim t_{\rm SE,J_2} \sim 10$ Myr, which roughly divides EKL-coupled and -decoupled planetary systems, as can be observed in Figure \ref{fig:figr013_EKL_disr}.
Consequently, in contrast to the top panels of Figure \ref{fig:r013_noEKL} for no stellar companion, the top panels of Figure \ref{fig:r013_EKL} show that systems with the circulating TTV phase and resonant angle of SE1 (max($|\Phi_{\rm TTV,1}|) \approx 180^\circ$ and max($\Delta \phi_1) \approx 360^\circ$), as well as disrupted SE pairs, can populate the lower-right quarter of the $m_{\rm CJ}$-$P_{\rm CJ}/P_{\rm SE1}$ parameter space.

In some cases, a cold Jupiter escapes or a stellar companion with initial $e_{\rm star2}\approx 1$ becomes temporarily unbound from the planetary system without disrupting resonance.
These violent outcomes are marked as $CJ$ (ejected cold Jupiter) and $S$ (quasi-unstable companion star) in Figure \ref{fig:r013_EKL}, using the same color code for max($|\Phi_{\rm TTV,1}|$) and max($\Delta \phi_1$).\footnote{min($q_{\rm CJ}$) is undefined when a cold Jupiter escapes, so we set min($q_{\rm CJ}$)=0 to display these cases in the middle and bottom panels of Figure \ref{fig:r013_EKL}.} They are rare in the r0.13 model: 2\% for quasi-unstable companion stars and 2.5\% for CJ ejections.

\begin{figure}[b]
    \centering
    \begin{minipage}{0.23\textwidth}
        \centering
        \includegraphics[width=\linewidth]{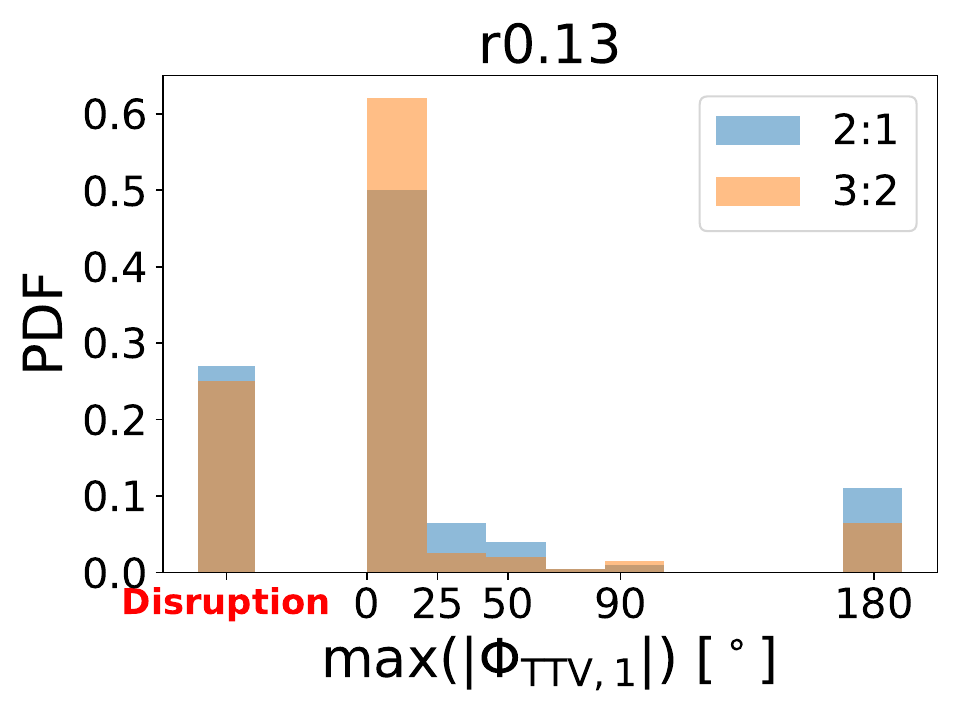}
    \end{minipage}
    \hfill
    \begin{minipage}{0.23\textwidth}
        \centering
        \includegraphics[width=\linewidth]{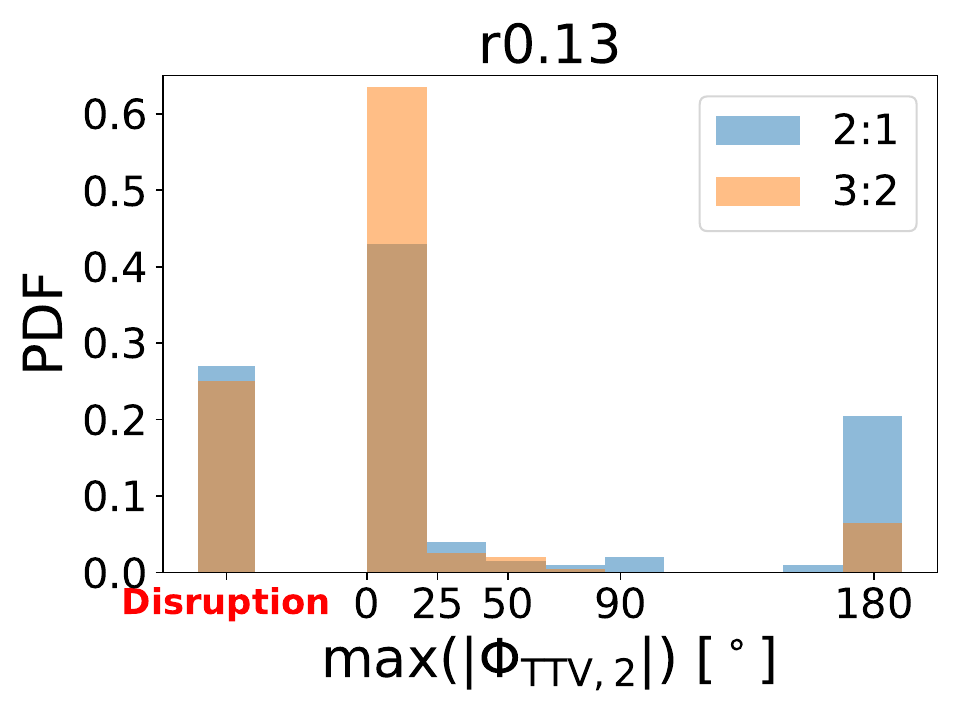}
    \end{minipage}

    \vspace{0 cm} 

    \begin{minipage}{0.23\textwidth}
        \centering
        \includegraphics[width=\linewidth]{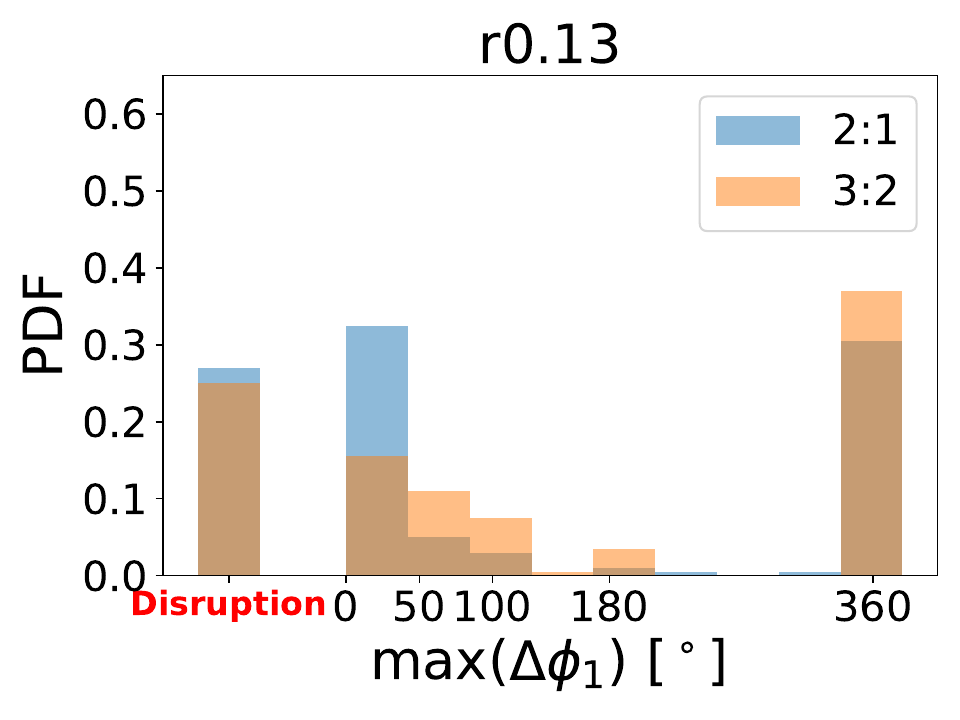}
    \end{minipage}
    \hfill
    \begin{minipage}{0.23\textwidth}
        \centering
        \includegraphics[width=\linewidth]{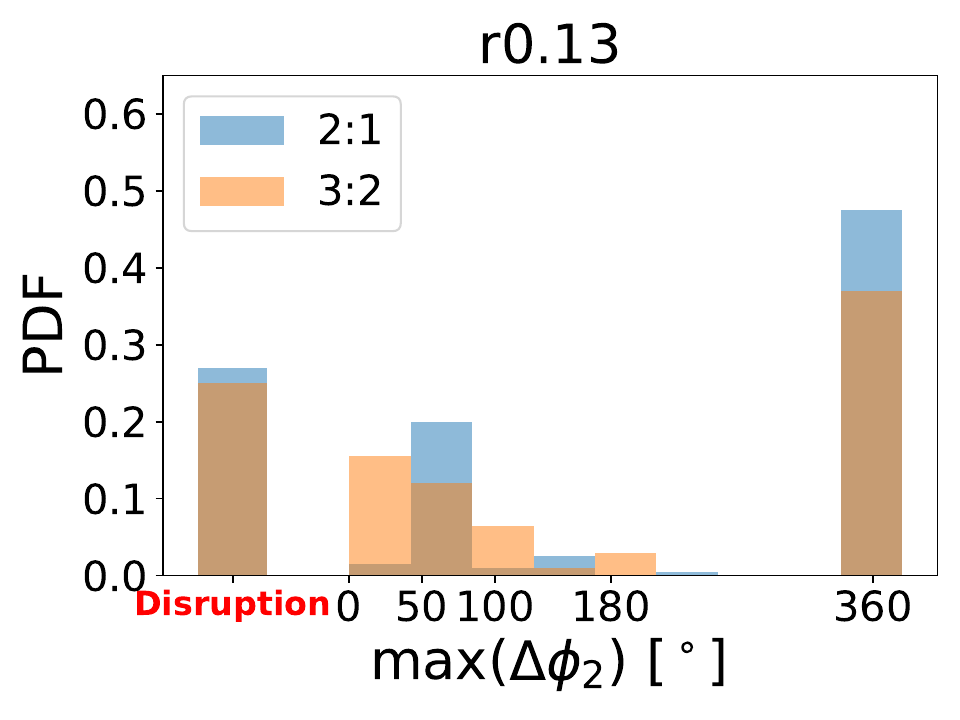}
    \end{minipage}
    
    \caption{Same as Figure \ref{fig:histr013_noEKL}, but in the presence of a stellar companion. In contrast to Figure  \ref{fig:histr013_noEKL}, a new population known as resonance disruption appears in these runs due to high eccentricity excitation by extreme EKL.
There are more systems having max($|\Phi_{\rm TTV}|$)$ > 25^\circ$ in the 2:1 resonance than those in the 3:2 one.}
    \label{fig:histr013_EKL}
\end{figure}

Figure \ref{fig:histr013_EKL} compares the PDF of the maximum TTV phase and maximum libration amplitude of resonant angles between the 2:1 and 3:2 resonances. Owing to the weaker resonant coupling in the 2:1 resonance than 3:2, the inner pair member is less affected by the external EKL perturbation, and therefore tends to have a smaller max($|\Phi_{\rm TTV}|$) and max($\Delta \phi$) than the outer pair member. As shown in the top panels of Figure \ref{fig:histr013_EKL}, $\sim 20$\% of SE2 have max($|\Phi_{\rm TTV,2}|)\approx 180^\circ$, while  $\sim 10$\% of SE1 have max($|\Phi_{\rm TTV,1}|)\approx 180^\circ$. In contrast, these possibilities drop to $\sim$6.5\%, and are almost identical for SE1 and SE2 in the 3:2 resonance. Interestingly, the most striking feature in the top panels of Figure \ref{fig:histr013_EKL} is the paucity of the maximum values of TTV phases between $\sim 90^\circ$ and $180^\circ$.

This dynamical imbalance associated with the 2:1 resonance becomes much more striking for the libration amplitude of resonant angles,  as shown in the bottom panels of Figure \ref{fig:histr013_EKL}. Many more 2:1 pairs have max($|\Phi_{\rm TTV,2}|$)$\approx 180^\circ$ than those with max($|\Phi_{\rm TTV,1}|$)$\approx 180^\circ$, while the 3:2 pairs have almost equal numbers for inner and outer circulating resonant angles. This leads to more (fewer) pairs with a circulating outer (inner) resonant angle in the 2:1 resonance than in the 3:2 resonance. Combining the 2:1 and 3:2 pairs, Figure \ref{fig:histr013_EKL} shows that $\sim$30-50\% of the EKL-perturbed pairs can reach the stage of circulating resonances. Libration amplitudes of resonant angles are more easily excited than the corresponding TTV phases (see Appendix \ref{sec:appendix}).

\begin{deluxetable*}{lcccc}
\tablecaption{Probability summary of circulating TTV phases and resonant angles in our three models for the 2:1 and 3:2 MMR SE pairs. The ``r0.13" model is initialized with a Rayleigh distribution of CJ eccentricities characterized by a mean value of 0.13, whereas the ``circ" model is initialized with CJs having $e_{\rm CJ}=0.01$, corresponding to nearly circular orbits.``w/o EKL" refers to the model with no stellar companion to excite the EKL mechanism. \label{table}}
\tablehead{
\colhead{model, MMR type} & \colhead{$\max (|\Phi_{\rm TTV,1}|)=180^\circ$} &  \colhead{$\max (|\Phi_{\rm TTV,2}|)=180^\circ$}  & \colhead{$\max (\Delta \phi_1)=360^\circ$} &  \colhead{$\max (\Delta \phi_2)=360^\circ$} 
}
\startdata
r0.13, 2:1 & 11\%  & 20.5\% &  30.5\% & 47.5\% \\
r0.13, 3:2 & 6.5\% & 6.5\% & 37\% & 37\% \\
r0.13 w/o EKL , 2:1 & 4\%  & 12.5\%  & 17\% & 33\% \\
r0.13 w/o EKL , 3:2 & 1\% & 1\% & 26\% & 26\% \\
circ, 2:1 & 3.5\% & 5.5\% & 12\% & 28\% \\
circ, 3:2 & 3.5\% & 3.5\% & 24.4\% & 24.4\% \\
\enddata
\end{deluxetable*}

Other than exciting larger max($|\Phi_{TTV}|$, the EKL mechanism induces disruption of MMRs, as has been shown in Figure \ref{fig:figr013_EKL_disr}.
Quantitatively, from the PDFs in Figure \ref{fig:histr013_EKL}, $\approx$25\% of the SE pairs experience resonant disruption. This disruption fraction is nearly identical for the 2:1 and 3:2 resonances, because the extreme EKL-driven eccentricity oscillations—arising from the large mutual inclination between the cold Jupiter and the companion star—are sufficiently destructive that the resonance strength does not play any significant dynamical role (see Figure \ref{fig:figr013_EKL_disr}). Consequently, the EKL mechanism renders planetary systems substantially more dynamically unstable and violent. 

\subsection{Compared to the circ model}
In addition to studying the r0.13 runs as our fiducial model, we also investigate the circ model from \citet{2025ApJ...980L..31W}, in which a cold Jupiter initially moves on a nearly circular orbit with $e_{\rm CJ}=0.01$. The circ model has long been considered, prior to Weldon et al., to disfavor stellar EKL as the primary driver of the eccentricities of cold Jupiters \citep{2005ApJ...627.1001T,2015ApJ...799...27P}. Therefore, it is instructive to supplement a study of this model and contrast the corresponding results with those from the r0.13 runs. 

\begin{figure}[ht]
    \centering
    \begin{minipage}{0.23\textwidth}
        \centering
        \includegraphics[width=\linewidth]{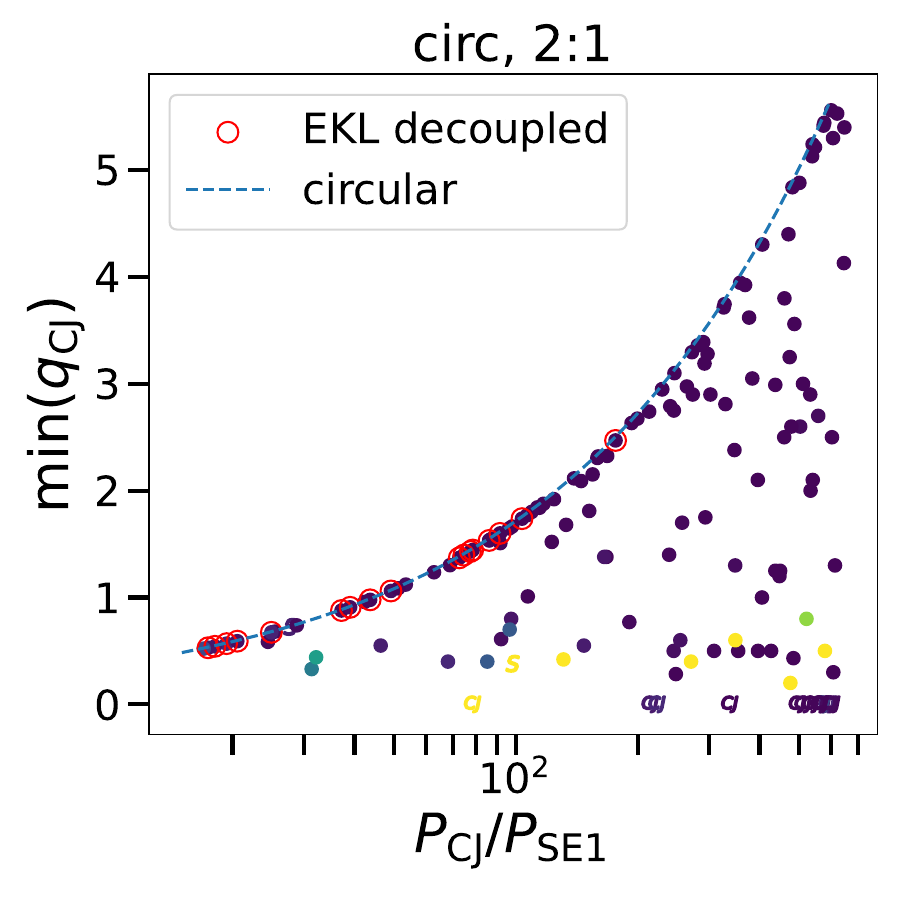}
    \end{minipage}
    \hspace{-0.017\textwidth}
    \begin{minipage}{0.23\textwidth}
        \centering
        \includegraphics[width=\linewidth]{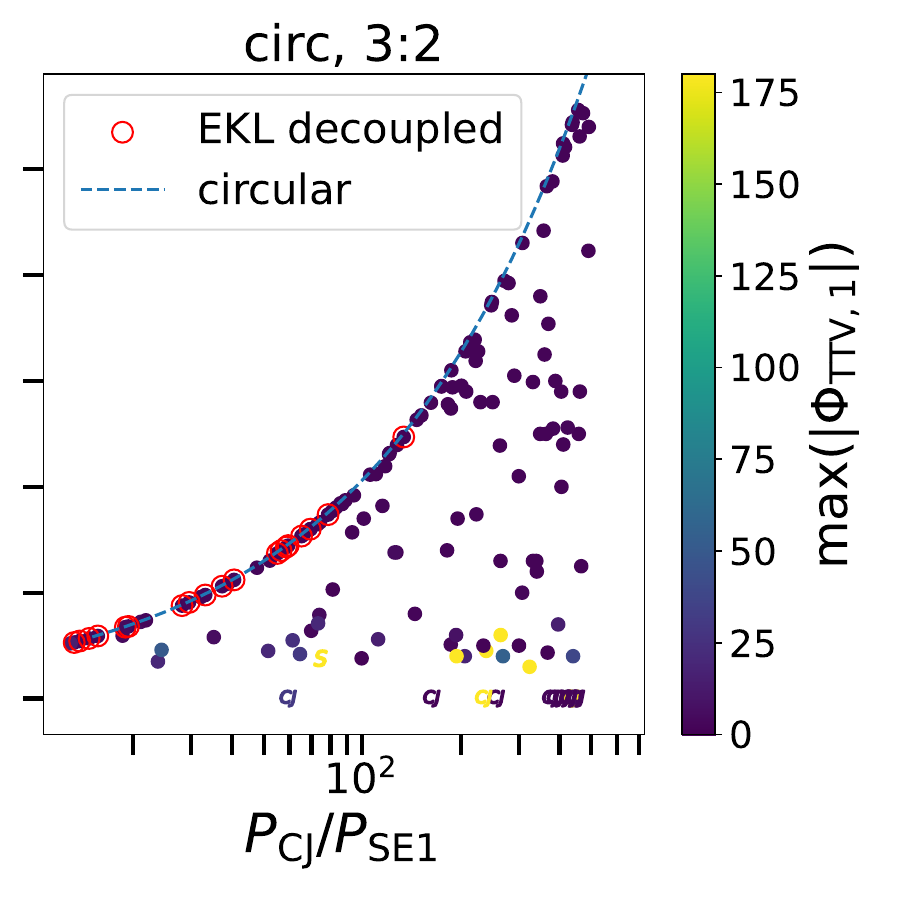}
    \end{minipage}

    \vspace{0 em} 

    \begin{minipage}{0.23\textwidth}
        \centering
        \includegraphics[width=\linewidth]{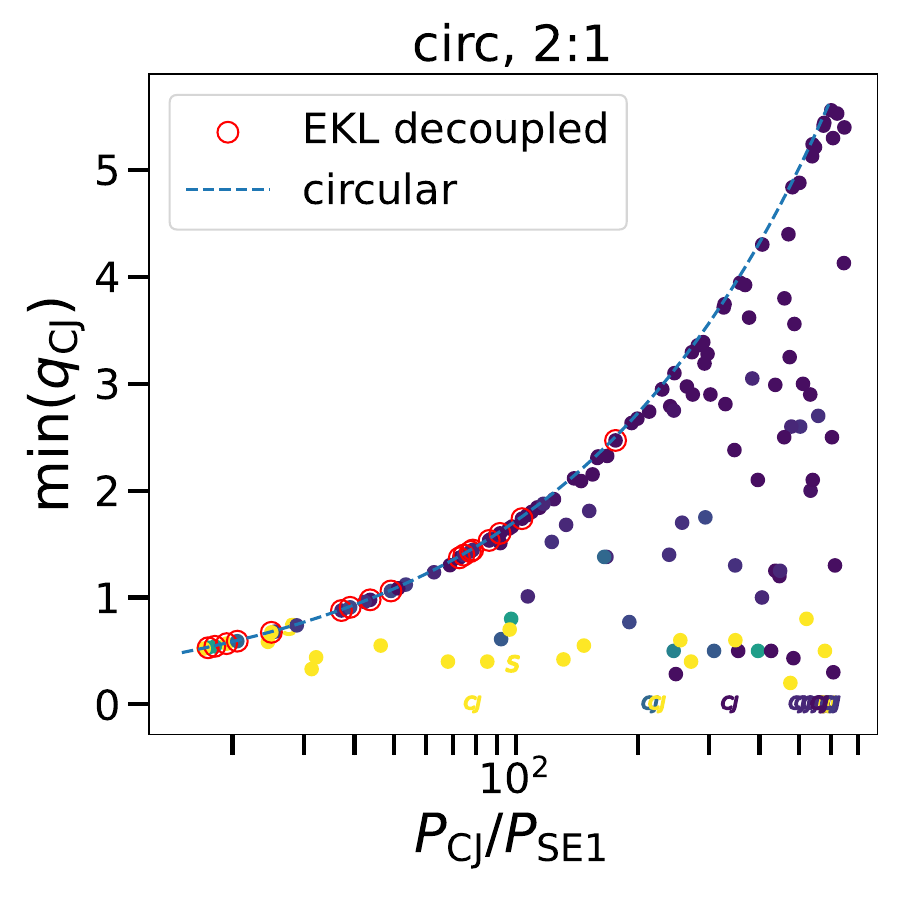}
    \end{minipage}
    \hspace{-0.017\textwidth}
    \begin{minipage}{0.23\textwidth}
        \centering
        \includegraphics[width=\linewidth]{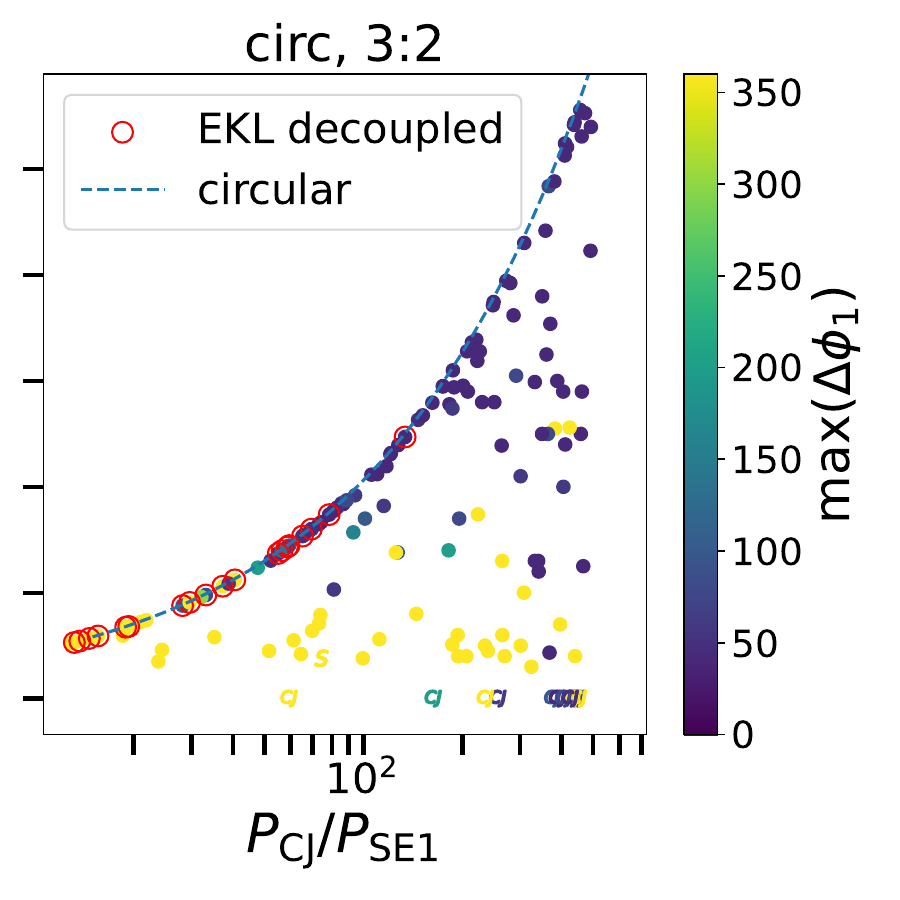}
    \end{minipage}
    \caption{Same as the middle and bottom panels of Figure \ref{fig:r013_EKL}, plotting the distributions of max($|\Phi_{\rm TTV,1}|$) and max($|\Phi_{\rm TTV,1}|$), but for the circ model.  The data point indicated by the letters $CJ$ and $S$, instead of a solid circle, represents an ejected cold-Jupiter and a quasi-unbound companion-star from the individual systems, respectively.\footnote{Please see the electronic version to read them clearly.} More systems with a cold Jupiter on a nearly circular orbit in the circ model than in the r0.13 model, as expected from \citet{2025ApJ...980L..31W}.}    
    \label{fig:circ}
\end{figure}

Figure \ref{fig:circ} shows the maximum TTV phase, max($|\Phi_{\mathrm{TTV},1}|$), and maximum libration amplitude, max($\Delta \phi_1$), for SE1 as functions of min($q_{\mathrm{CJ}}$) and $P_{\mathrm{CJ}}/P_{\mathrm{SE1}}$ in the circ model. Unlike the middle and bottom panels of Figure \ref{fig:r013_EKL} for the r0.13 simulations, many more systems lie near the $e_{\mathrm{CJ}}=0$ locus, indicating that numerous configurations host a cold Jupiter on an approximately circular orbit, even though EKL-driven systems still span min($q_{\mathrm{CJ}}$) values below the curve denoted by ``circular" ($e_{\mathrm{CJ}}=0$). This matches the excess of low-$e_{\mathrm{CJ}}$ outcomes in the circ model reported by Weldon et al. As before, smaller min($q_{\mathrm{CJ}}$) values preferentially yield larger max($|\Phi_{\mathrm{TTV},1}|$) and max($\Delta \phi_1$) for resonant pairs, due to closer proximity and thus stronger secular perturbations.

Inspection of the color-coded points in Figure \ref{fig:circ} shows that more systems with 2:1 resonant pairs reach higher max($\Phi_{\mathrm{TTV},1}$) than those with 3:2 pairs, whereas fewer 2:1 systems attain higher max($\Delta \phi_1$) than 3:2 systems. As explained previously, this results from the asymmetric effect of the indirect potential in 2:1 MMR pairs, which weakens the resonant coupling that relays the external secular EKL perturbation from SE2 to SE1.

\begin{figure}[ht]
    \centering
    \begin{minipage}{0.23\textwidth}
        \centering
        \includegraphics[width=\linewidth]{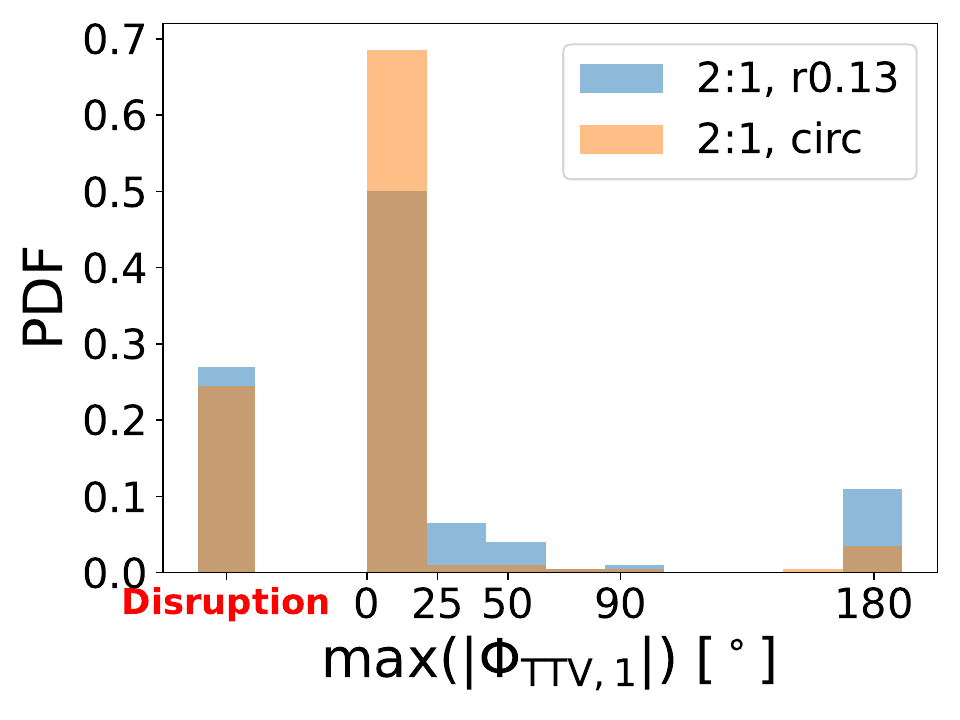}
    \end{minipage}
    \hfill
    \begin{minipage}{0.23\textwidth}
        \centering
        \includegraphics[width=\linewidth]{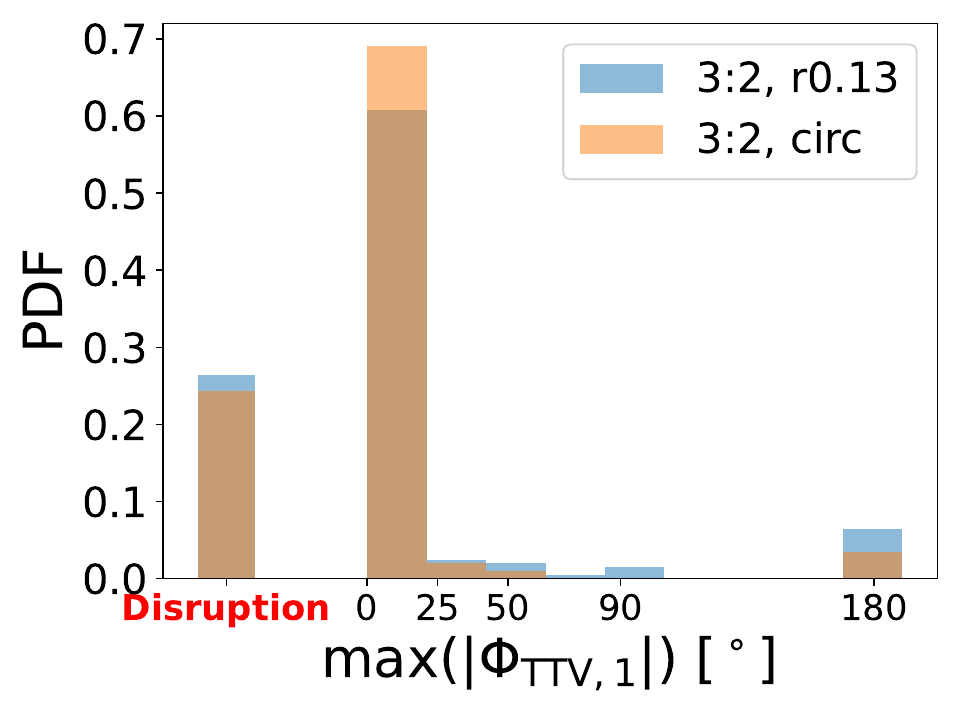}
    \end{minipage}

    \vspace{0 em} 

    \begin{minipage}{0.23\textwidth}
        \centering
        \includegraphics[width=\linewidth]{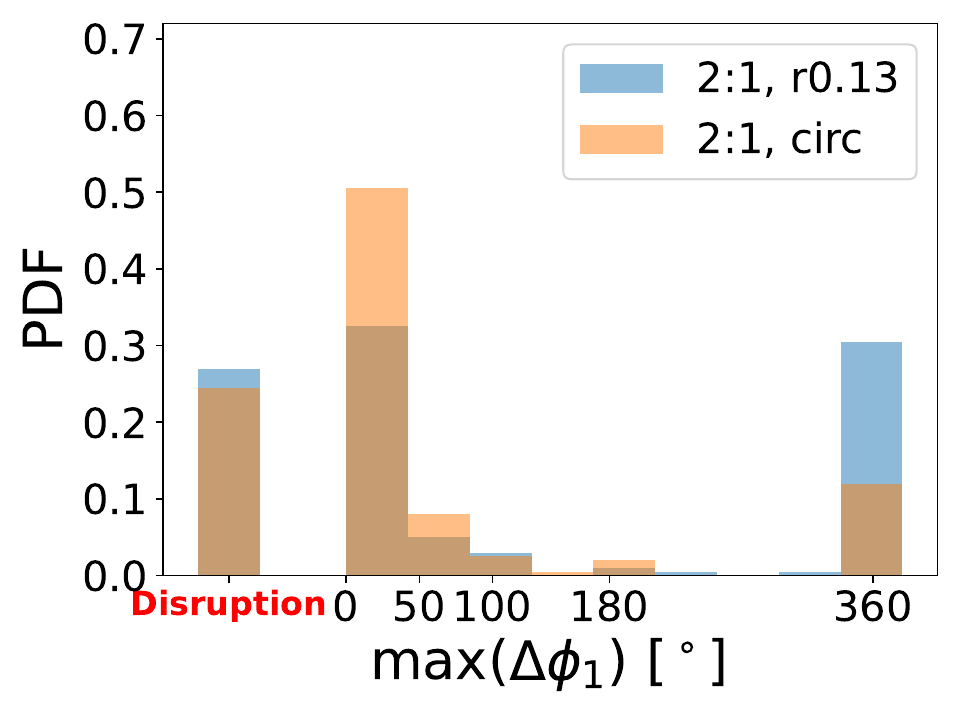}
    \end{minipage}
    \hfill
    \begin{minipage}{0.23\textwidth}
        \centering
        \includegraphics[width=\linewidth]{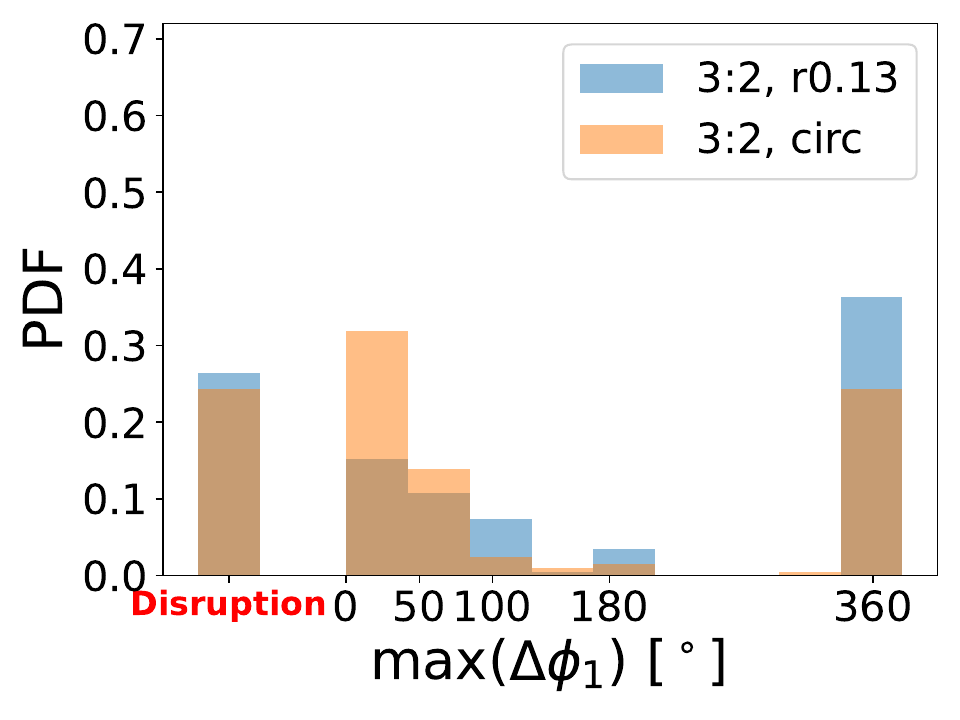}
    \end{minipage}
    \caption{A comparison of the PDF histograms of max($|\Phi_{\rm TTV,1}|$) and max($\Delta \phi_1$) between the r0.13 and circ models. The two left panels display results for systems in a 2:1 pair, whereas the two right panels correspond to systems in a 3:2 resonance. Using max($|\Phi_{\rm TTV,1}|$) and max($\Delta \phi_1$) as dynamical activity indicators for resonant pairs, the r0.13 model exhibits systematically higher levels of dynamical excitation than the circ model. }    
    \label{fig:histcirc}
\end{figure}

The histograms in Figure \ref{fig:histcirc} show the PDFs of max($|\Phi_{\rm TTV,1}|$) and max($\Delta \phi_1$) for SE1, allowing quantitative model comparison. For both the 2:1 (2 left panels) and 3:2 (2 right panels) resonant pairs, a larger fraction of systems reach higher values of max($|\Phi_{\rm TTV,1}|$) and max($\Delta \phi_1$) in the r0.13 model than in the circ model. Specifically, $\approx 20$\% of 2:1 pairs and $\approx 10$\% of 3:2 pairs with max($|\Phi_{\rm TTV,1}|) \lesssim 20^\circ$ in the circ model are excited to higher values in r0.13, and $\approx 20$\% of 2:1 and 3:2 pairs with max($\Delta \phi_1) \lesssim 40^\circ$ in the circ model are similarly excited. Thus, systems with a Rayleigh distribution of $e_{\rm CJ}$ (mean 0.13) are dynamically hotter than those with a nearly circular cold Jupiter, and the circ model excites large TTV phases and resonant libration amplitudes less efficiently, by $\sim 20$\%.

Figure \ref{fig:histcirc} shows that 2:1 pairs more often reach larger max($|\Phi_{\rm TTV,1}|$) than 3:2 pairs, consistent with weaker resonant coupling in the 2:1 resonance, while 3:2 pairs more often reach larger max($\Delta \phi_1$). As explained earlier, stronger coupling in the 3:2 case more efficiently relays the EKL perturbation and excites the libration amplitude of the inner pair member than in the 2:1 case. 

In addition to the PDFs in Figure \ref{fig:histcirc}, Table \ref{table} summarizes the probabilities of circulating TTV phases and resonant angles from our three statistical models, including the r0.13 model without a stellar companion. The table also lists the corresponding SE2 results to distinguish coupling strengths for different MMRs (also see Figures \ref{fig:histr013_noEKL} and \ref{fig:histr013_EKL}). For 3:2 resonant pairs, the circulating probabilities of $\Phi_{\mathrm{TTV}}$ and $\phi$ are nearly identical between SE1 and SE2. In contrast, for 2:1 resonant pairs, these probabilities are consistently lower for SE1 than for SE2 in all models, indicating weaker resonant coupling in the 2:1 case. As previously discussed and elaborated by Table \ref{table}, the r0.13 model with stellar EKL produces a higher fraction of circulating $\Phi_{\mathrm{TTV}}$ and $\phi$ than either the r0.13 model without EKL or the circ model.

Comparing Figure \ref{fig:circ} with Figure \ref{fig:r013_EKL} shows that a larger fraction of cold Jupiters (labeled as ``$CJ$" in the plots) are ejected from systems that retain SE pairs in the circ model than in the r0.13 model. 
To investigate this difference, we evaluate the orbital instability criterion for the initial cold Jupiter and stellar companion orbits in our simulations \citep{2001MNRAS.321..398M}:
\begin{eqnarray}
\frac{a_{\rm star2}}{a_{\rm CJ}} < && 2.8 \left( 1 + \frac{m_{\rm star2}}{m_*+m_{\rm CJ}} \right)^{2/5} \frac{(1+e_{\rm star2})^{2/5}}{(1-e_{\rm star2})^{6/5}} \nonumber \\
 &&\times \left(1- \frac{0.3\, i_{\rm star2}}{180^\circ} \right).
\end{eqnarray}
We find that $\approx 11$\% of planetary systems in both the r0.13 and circ models are formally unstable by this criterion. In the circ model, most of these unstable systems eject the cold Jupiter, 
while retaining the SE pairs. In the r0.13 model, many unstable systems not only lose the cold Jupiter but also disrupt the SE pairs, so they do not appear in the middle and bottom panels of Figure \ref{fig:r013_EKL}, which show only surviving SE pairs. 
Because cold Jupiters in the r0.13 model typically reach higher eccentricities, those perturbed by a highly eccentric or close stellar companion are more likely to disrupt SE pairs than in the circ model. The impact on the PDFs is small, however: Figure \ref{fig:histcirc} shows that the SE-pair disruption probability in the r0.13 model is only slightly higher than in the circ model.

Thus, the fraction of systems undergoing resonance disruption remains similar in the circ simulations and the r0.13 model ($\approx 25$\%). Although the circ model overproduces low-$e_{\rm CJ}$ cold Jupiters and underproduces those at moderate $e_{\rm CJ}$, it leaves the high-$e_{\rm CJ}$ cold Jupiter population largely unchanged relative to r0.13, consistent with \citet{2025ApJ...980L..31W}. As noted earlier, it is the systems with high $e_{\rm CJ}$, driven by extreme EKL excitation, that primarily undergo resonance disruption (see Figure \ref{fig:figr013_EKL_disr}).

\subsection{Resonance disruption and long-term evolutions}
\label{sec:longterm}

Thus far, we have performed N-body simulations for 16 Myr. As a result, resonant pairs are directly disrupted by extreme EKL oscillations, in which the cold Jupiter plunges significantly toward the pair and disrupts it on this timescale. Although the stellar EKL modulates the orbital evolution of a cold Jupiter on $\sim$ Myr-Gyr timescales in the r0.13 model without the internal SE pairs \citep{2025ApJ...980L..31W}, Figure \ref{fig:figr013_EKL_disr} and Section \ref{sec:r0.13} show that systems with $t_{\rm EKL} \gtrsim 10$ Myr are dominated by closer-in cold Jupiters that are effectively decoupled from stellar EKL due to the J$_2$ potential from the SE pair. This implies that our numerical analyses, which run for only 16 Myr, provide an informative description of resonance excitation and disruption caused by stellar EKL.

To further verify our expectations, we run simulations for the r0.13 model up to $\approx$ 100 Myr. We find that 6\% more of the 2:1 pairs and 4.5\% more of the 3:2 pairs are disrupted in our sample. However, almost all of them are not disrupted by the EKL with $t_{\rm EKL}>16$ Myr but are disrupted by their excited TTV phases, by extension, their free eccentricities. Specifically, most new pairs that disrupt during the extended time are those surviving systems with modest and large max($\Phi_{\rm TTV,1}$) around the pair-disruption population (marked by grey crosses) in Figure \ref{fig:figr013_EKL_disr}. In conclusion, there are two types of EKL-induced resonance disruption, associated with short- and long-timescale processes. A fraction of resonant pairs are disrupted directly by extreme EKL within $\sim 10$ Myr ($\sim$ 25\%), and subsequently the surviving pairs with EKL-induced high free eccentricities would disrupt on a longer timescale in the r0.13 model.

\section{Discussions}
\label{sec:discussion}

\subsection{asymmetric TTV correlation in near 2:1 pairs}
\label{sec:asymTTV}

TTV signals vary with time. Analytical TTV theory indicates that when the linear combination of the planets’ free eccentricities for a near-resonant pair, $|\mathcal{Z}_{\rm free}|$, evolves to exceed $\sim \Delta$, the corresponding TTV phases, $\Phi_{\rm TTV}$, likewise become large \citep[][see also Appendix \ref{sec:appendix}]{2012ApJ...761..122L}.
When this happens under the secular perturbation of a third body, $|\mathcal{Z}_{\rm free}/f_1| \approx \mathcal{D} \equiv |e_{\rm SE1}-|f_2/f_1|e_{\rm SE2}|$, where the order-of-unity coefficients $f_1$ and $f_2$ describe the strength of resonant interactions of the pair \citep{1999ssd..book.....M}. \citet{2023MNRAS.522.1914C} used max($\mathcal{D}/\Delta$) as a proxy for TTV-phase excitation.

\begin{figure}[ht!]
\plotone{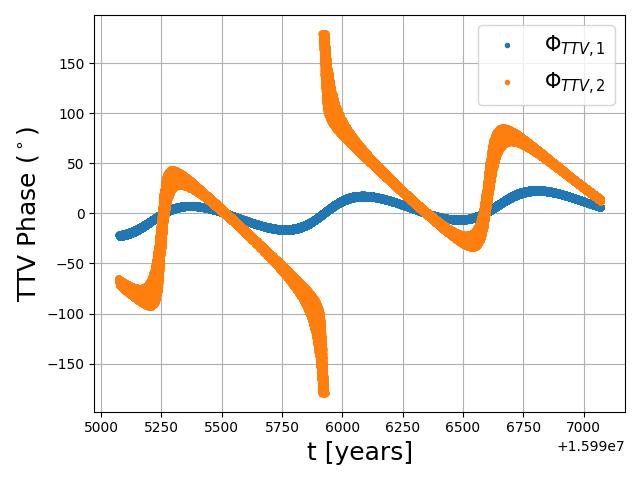}
\caption{A short segment of the TTV-phase evolution in one of the r0.13 runs for a 2:1 pair, showing an asymmetric correlation between $\Phi_{\rm TTV,1}$ and $\Phi_{\rm TTV,2}$. 
\label{fig:asymTTV}}
\end{figure}

The resonant pair with modest max($\Phi_{\rm TTV,1}$) can sometimes pass through a stage in which $\mathcal{D}/\Delta \sim 1$, thereby exhibiting asymmetric TTV signals (see Appendix \ref{sec:appendix}). The condition $\mathcal{D}/\Delta \sim 1$ often occurs in the ensemble of our planetary systems in the r0.13 model, which covers a broad range of orbital distributions and thus so does $\mathcal{D}/\Delta$.
Figure \ref{fig:asymTTV} illustrates a temporal segment of an example for a 2:1 pair with max($\mathcal{D}/\Delta) \approx 0.7$, showing that $|\Phi_{\rm TTV,2}| > |\Phi_{\rm TTV,1}|$ most of the time. The difference in TTV phases can sometimes be large during the evolution. The same phenomenon can also be noticed in the examples illustrated in Figure \ref{fig:r013_EKLexample} for a long period of time. 

\begin{figure}[ht!]
    \centering
    \plotone{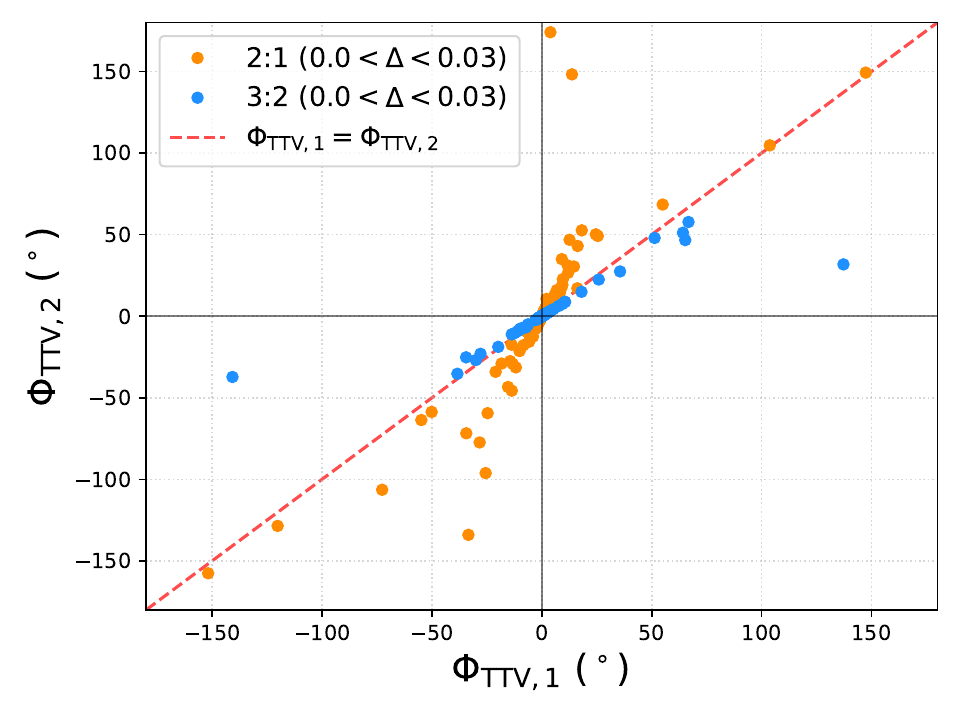}
    \vspace{0.2cm}
    \plotone{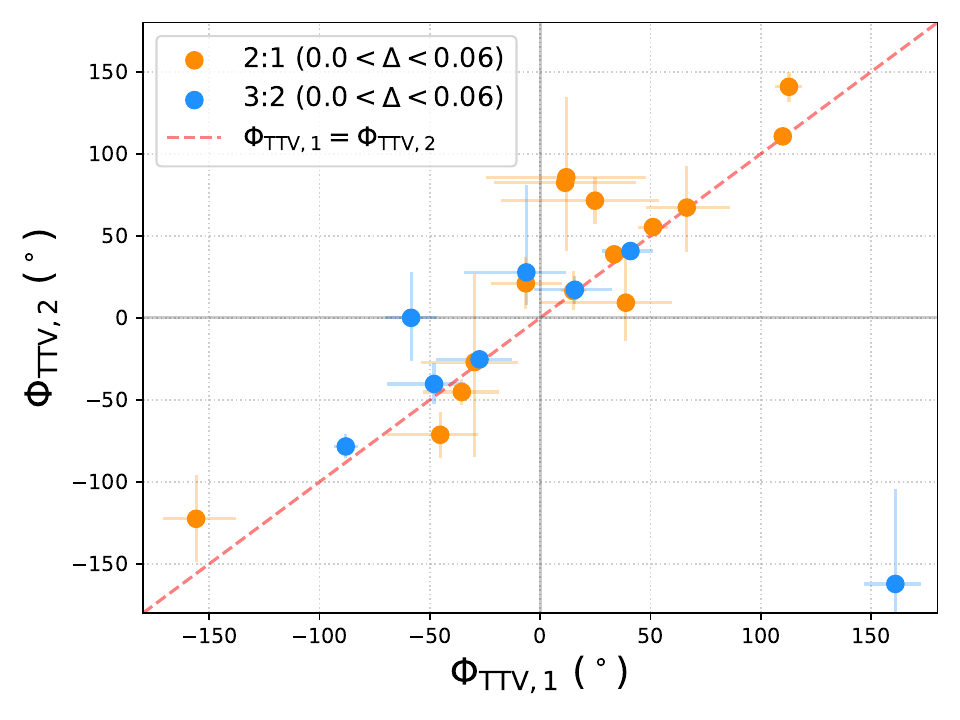}
    \caption{The upper panel shows a distribution of the simulated TTV phases of the 2:1 and 3:2 resonant pairs at $t=100$ Myr for the r0.13 model.
The lower panel shows the observed Kepler TTV phases of the 2:1 and 3:2 resonant pairs, taken from \citet{2014ApJ...787...80H}, for the inner and outer members of each pair. $\Phi_{\rm TTV,2}$ has been shifted by $180^\circ$ to conform to the sign convention adopted by \citet{2024ApJ...971....5W} and employed throughout this work. Kepler pairs with uncertainties greater than 60$^\circ$ have been removed. The error bars indicate 1$\sigma$ uncertainties. The values of $\Delta$ for the data points span the interval from 0 to 0.03 in the upper panel and from 0 to 0.06 in the lower panel. The data points are color-coded to distinguish systems in 2:1 and 3:2 MMRs.The red dashed line corresponds to the exact symmetric correlation of TTV signals, i.e., $\Phi_{\rm TTV,1} = \Phi_{\rm TTV,2}$.
    \label{fig:TTVcomp}}
\end{figure}

The top panel of Figure \ref{fig:TTVcomp} shows the instantaneous distribution of $\Phi_{\rm TTV,1}$ vs. $\Phi_{\rm TTV,2}$ of both near 2:1 and 3:2 surviving pairs at the end of our numerical integration, i.e., $t=100$ Myrs. in the r0.13 model. 
Many of the 2:1 pairs are distributed away from the exact symmetric correlation of TTV signals (red dashed line) with $|\Phi_{\rm TTV,1}| > |\Phi_{\rm TTV,2}|$, which arise from $\mathcal{D}/\Delta \sim 1$ at this time. In comparison, some 3:2 pairs tend to have $|\Phi_{\rm TTV,1}| < |\Phi_{\rm TTV,2}|$ with much smaller deviations from the exact symmetric TTV correlation as a result of stronger dynamical coupling between pair members. Furthermore, most data points are distributed within $|\Phi_{\rm TTV}|\lesssim 20^\circ$, consistent with the PDF of the maximum TTV phases shown in Figure \ref{fig:histr013_EKL}. $\Delta$ of most of the simulated SE pairs remains $<0.03$ due to our initial setup (see Section \ref{sec:setup}).\footnote{Hence, the numerically simulated distribution of TTV phases remains nearly unchanged across the extended interval \(0 < \Delta < 0.06\) used by \citet{2024ApJ...971....5W} (see below).}

The lower panel of Figure \ref{fig:TTVcomp} presents the Kepler TTV phase distributions to compare with our numerically generated distribution. This observed distribution is nearly identical to Figure 2 of \citet{2024ApJ...971....5W} for the Kepler TTV data from \citet{2014ApJ...787...80H} with $0<\Delta<0.06$, but here each data point is color-coded to distinguish 2:1 and 3:2 MMRs. 
Although the Kepler sample size is small and the associated observational uncertainties are significant, the observed distribution suggests a tendency toward a greater excess of asymmetric 2:1 systems lying above (below) the dashed line in the first (third) quadrant of the TTV phase diagram. This behavior may point to a qualitative correspondence between the observed Kepler 2:1 MMRs and the theoretically predicted asymmetric TTV phase distributions driven by the indirect potential in our numerical model. 

Nonetheless, in Appendix \ref{sec:KeplerTTV_comp}, we show that when restricting the sample to systems with $0 < \Delta < 0.03$, similar to the simulated range of $\Delta$, the Kepler TTV phase distribution appears less scattered from the red dashed line and thus is more symmetric than that obtained from our numerical model. This may imply that the SE pairs in the observations are, on average, more dynamically excited than the pairs produced in our EKL-based simulations.

Combining the 2:1 and 3:2 pairs without distinguishing $\Phi_{\rm TTV,1}$ and $\Phi_{\rm TTV,2}$ in the fiducial r0.13 model, the top panels of Figure \ref{fig:histr013_EKL} imply that $\approx$ 26\% of the surviving pairs can reach max($\Phi_{\rm TTV}) \gtrsim 20^\circ$. Inspecting the results in more detail, we find that $\approx$ 35\% of the surviving pairs can attain max($\Phi_{\rm TTV}) \gtrsim 10^\circ$. Because most of these {\it maximum} TTV phases are either $< 90^\circ$ or $\sim 180^\circ$, EKL-induced TTV phases tend to have more small, non-zero values, as suggested by Kepler TTV data in the lower panel of Figure \ref{fig:TTVcomp}. Although in our model the TTV phases are mostly confined to max($\Phi_{\rm TTV}) \lesssim 20^\circ$ and appear to be more asymmetrically correlated for 2:1 pairs than is seen in the Kepler data, we note that our model follows the r0.13 run from \citet{2025ApJ...980L..31W}, with an initial Rayleigh distribution of the CJ eccentricity. The formation of this Rayleigh distribution, say, due to giant-planet scattering, could have excited the libration amplitude of resonant angles to some extent \citep{2026ApJ..1003...45G}, such that the subsequent (or simultaneous) EKL perturbations would excite them more than what we have produced in this work. Furthermore, in this work, the SE pair lies at $\lesssim 0.127$ au, deep in the potential well of the host star of one solar mass. Apart from that, we do not include resonant chains for simplicity in this study. Therefore, our EKL-induced TTV results are encouraging, and more comprehensive numerical studies would further clarify them.

\subsection{Other orbital resonance configurations}

We restrict our analysis to SE pairs near the 2:1 and 3:2 MMRs, because these are the dominant resonance configurations observed among Kepler SE pairs, and their TTV phases have been extensively characterized in the literature.
Some SE pairs instead reside in more compact MMRs, such as the 4:3 and 7:5 resonances in the Kepler and TESS measurements. For planetary pairs in first-order MMRs with period ratios close to $(q+1)/q$, the mutual orbital separation decreases with increasing $q$, whereas the resonance width increases. Consequently, the resonance coupling is weakened at larger $q$ as a result of enhanced resonance overlap \citep[e.g.,][]{2013ApJ...774..129D,2018AJ....156...95H}. 
In the case of second-order MMRs, the strength of the resonant coupling between planets is reduced relative to that of first-order resonances by a factor proportional to the orbital eccentricity \citep[e.g.,][]{1999ssd..book.....M,2024AJ....168..239D,2025AJ....169..323L}. We therefore expect that the libration amplitude of resonant angles and corresponding TTV phases associated with the 4:3 and 7:5 commensurabilities are more easily excited by our EKL model than those for the 3:2 and 2:1 resonant pairs.

Many of the resonant SE pairs are embedded within resonant chains.
In general, resonant chains are more dynamically fragile and more susceptible to disruption than isolated resonant pairs because of richer dynamics. Under appropriate conditions, resonant chains can facilitate resonance overlap and the development of secondary resonances, which can lead to the growth of libration amplitude and dynamical instability on their own \citep[e.g.,][]{2012Icar..221..624M,
2022Icar..38815206G,2025AJ....169..323L}. In this regard, the TTV phases and resonant libration amplitudes of SE pairs in a resonant chain may be more readily excited by a CJ that is perturbed through EKL interactions with a stellar companion. Conversely, depending on the orbital spacing and masses of the SEs, resonant chains may enhance the precession rate of the CJ, thereby suppressing the EKL mechanism and, in turn, diminishing the external excitation of resonant dynamical properties. A detailed investigation of these effects is deferred to future work.

\subsection{Influence of stellar and SE masses}
\label{sec:stellar_mass}
In this study, our fiducial configuration is a binary system of two 1 $M_\odot$ stars and a SE pair of two 8 $M_\oplus$ planets, applied to an ensemble of 200 numerically simulated planetary systems. Owing to the limited sample size, our approach of using a single stellar mass differs from that of \citet{2025ApJ...980L..31W}, who considered a Salpeter power-law distribution for stellar masses between 0.6 and 1.6 $M_\odot$ in a larger sample to include one more dimension of the input stellar parameter for the EKL effect.\footnote{For example, \citet{2025ApJ...980L..31W} simulated 800 planetary systems for the r0.13 model.} The EKL efficiency arises from the competition between the EKL oscillation and the precession of the CJ induced by the inner pair. Because stars with mass 0.6 $M_\odot$ are more common in the Salpeter distribution, we adopt this mass to evaluate the possible different outcome due to stellar masses. Equations (\ref{eq:t_EKL}) and (\ref{eq:t_J2}) indicate that $t_{\rm EKL}$ and $t_{\rm SE,J_2}$ scale as $m_*^{-1/2}$, where we consider $m_*=m_{\rm star2}$ to maximize the effect of stellar masses. Thus, $0.6^{-1/2} = 1.3$. This is still close to 1 and therefore slightly extends the maximal timescale for the EKL action from $\sim 10$ to $\sim 13$ Myrs (see Section \ref{sec:r0.13}).  
Consequently, using the Salpeter power-law distribution is unlikely to introduce a substantial difference in our results running for 16-100 Myrs.

Regarding the mass of the SE, Equation (\ref{eq:t_J2}) indicates that the precession rate of a CJ driven by the inner pair scales linearly with the planetary mass. Consequently, a smaller SE mass enhances the stellar EKL effect on the CJ, thereby increasing the efficiency of TTV-phase excitation, and vice versa. However, even when allowing the SE mass to span the range $\sim$2–10 $M_\oplus$, the corresponding variation in the precession rate, by a factor of $8M_\oplus/2 M_\oplus \sim 4$ relative to our adopted mass, still appears insufficient to enable systems with long EKL cycles ($\gtrsim 100$ Myr) to excite the TTV phases.
In addition to the J$_2$ contribution, the mass ratio within the resonant SE pair also influences the excitation efficiency. In the mass range within the $\sim$2–10 $M_\oplus$ interval, the extreme mass ratio is $m_{\rm SE1}/m_{\rm SE2} \sim 2 M_\oplus/10 M_\oplus \sim 0.2$, and \citet{2023MNRAS.522.1914C} demonstrated that resonant pairs with mass ratios $<1$ are more susceptible to external secular perturbations over a certain region of parameter space (see their Fig. 10). As noted by \citet{2023MNRAS.522.1914C}, an equal-mass SE pair tends to yield conservatively low values of $\max(\mathcal{D}/\Delta)$. Accordingly, our statistical analysis likely underestimates the true probability of TTV-phase excitation induced by the stellar EKL mechanism.

\subsection{Cold-Jupiter formation in eccentric and close binaries}
\label{sec:CJformation}
Our study assumes a uniform distribution of initial binary eccentricity, as in \citet{2025ApJ...980L..31W}, without addressing its origin, especially its impact on planet formation. Observations reveal that stellar binaries with separations less than $\lesssim$ 100 au tend to have an eccentricity distribution consistent with uniform, while systems with wider separations exhibit thermal or even suprathermal distributions \citep[e.g.,][]{2010ApJS..190....1R, 2017ApJS..230...15M, 2022MNRAS.512.3383H, 2023ASPC..534..275O}. 
As \citet{2025ApJ...980L..31W} noted, using a uniform eccentricity distribution is more conservative than a thermal one for binary eccentricities \citep{1919MNRAS..79..408J}, leading to weaker EKL effects. The thermal distribution in wider binaries likely results from post-formation processes, such as the dissolution of their host stellar clusters \citep{2010MNRAS.404.1835K}. 

Conversely, a nearly uniform eccentricity distribution at smaller separations ($\lesssim 100$ au) is more naturally interpreted as a consequence of binary formation within a gaseous protoplanetary disk, rather than dynamical processing in a stellar cluster \citep[e.g.,][]{2022MNRAS.512.3383H}. Under this assumption, the presence of a close and eccentric stellar companion suggests a potential suppression of planet formation: the companion not only tidally truncates the circumstellar disk but also excites perturbations in planetesimals and planetary embryos, thereby inhibiting the formation of cold Jupiters (CJs). Potential tidally truncated circumstellar disks in widely separated young multiple-star systems were probed in a star-forming region using the Atacama Large Millimeter/submillimeter Array. \citep[e.g.,][]{2022A&A...662A.121R}.

Within the core accretion framework, specifically via pebble accretion in an S-type coplanar binary configuration, theoretical models indicate that a planet with a mass of $\sim 6\,M_{\rm Jup}$ can still form at radii as small as $\sim 0.3$ times the tidal truncation radius ($R_{\rm trunc}$), owing to the high efficiency and rapidity of core growth \citep{2026A&A...708A..38N}. In our r0.13 model, we find that the majority of CJs in stellar binaries with separations $< 100$ au orbit at distances exceeding $0.3\,R_{\rm trunc}$, which points to significant formation challenges for the CJs in our sample (see Appendix \ref{sec:truncation} for details). 
However, misalignment between the circumstellar disk and the binary orbital plane can reduce the effective tidal truncation torque, thereby weakening disk truncation and potentially alleviating these formation constraints \citep[e.g.,][]{2015ApJ...800...96L, 2015MNRAS.452.2396M}.

Alternatively, a close-in CJ in a tight binary system may originate through disk fragmentation triggered by gravitational instability at a comparatively wide initial separation within a massive protoplanetary disk, followed by inward migration \citep[e.g.,][]{2022MNRAS.512.3383H, 2026arXiv260302395Z}. Although it is often stated that orbits of disk and binary tend to be aligned in this scenario, the eccentricity and inclination distributions of stellar binaries formed via this disk-fragmentation pathway have not yet been systematically characterized. Furthermore, it remains uncertain whether the eccentricity distributions of CJs produced by these formation channels are well described by a Rayleigh distribution, which constitutes one of the key initial assumptions in our EKL analysis. A rigorous understanding of planet formation, binary evolution, and star cluster dynamics requires integrated theoretical modeling and observational constraints to refine physical models and statistical inferences.

\subsection{Formation of $\Delta \sim 0.01$}
\label{sec:formation_Delta}
While we study the excitation of TTV phases and libration amplitudes of resonant angles through a cold Jupiter's EKL interacting with an outer stellar companion, our model does not address the formation of $\Delta \sim$ a few percent for a librating near-resonant pair, but assumes it as an initial condition, as done by \citet{2023MNRAS.522.1914C}. In theory, the wider spacing of period commensurabilities like the 3:2 and 2:1 with librating resonant angles could be achieved by resonant repulsions during convergent migration in a gas-poor transitional disk \citep{2020MNRAS.495.4192C} or by revising classic Type-I migration for super-Earths to open particle gaps that allow wave-planet interactions in the horseshoe orbital regions \citep{2021ApJ...921..142C}. As mentioned in the Introduction, mechanisms associated with disk turbulence and divergent migrations in the post-disk epoch can potentially increase the TTV phases and the libration amplitudes of resonant angles at the early stages of planetary systems. In these scenarios, our EKL model might further excite free eccentricities and facilitate resonant disruption to a greater extent than we present in this study. A full assessment requires an extensive, careful study that incorporates various mechanisms, including short-range forces such as tidal and stellar quadrupole effects, at different evolutionary stages, and that considers a wider parameter space covering various SE masses, second-order MMRs, and planetary/stellar multiplicity \citep[e.g.,][]{2023MNRAS.522.1914C,2024AJ....167...55H,2025ApJ...980L..31W,2026ApJ...996...59Y}.

\subsection{Future prospect}

Our EKL model of resonant TTV phases relies on Kepler discoveries and the presence of a cold Jupiter and a distant stellar companion. Although we follow Weldon et al. in considering the observed distributions of initial conditions for these outer perturbers, the observations are limited by detection thresholds and selection effects -- the Kepler resonance sample is small and has significant uncertainties, and distant perturbers are difficult to detect using the radial velocity technique \citep[e.g.,][]{2024ApJ...962L..21K}.
These observational limitations will be relaxed in the era of TESS, Gaia, Rubin, and ET 2.0, enabling an extended transit-time baseline, detection of more TTV signals across younger to older planetary systems, and probing of super-Jupiters and substellar companions, providing updated information to greatly improve the probability distributions of input parameters in this work \citep[e.g.,][]{2018haex.bookE...7A,2022arXiv220606693G,2023arXiv231201903M,2026AJ....171...18L}.

\section{Summary and Conclusions}
Motivated by recent observational evidence that most young super-Earth systems are in near-MMRs, we perform N-body simulations to explore the excitation of TTV phases and the libration amplitude of resonant angles for 2:1 and 3:2 resonant pairs driven by a cold Jupiter interacting with a stellar companion via the EKL mechanism. For the statistical study, we adopt the r0.13 ensemble model of \citet{2025ApJ...980L..31W} as our fiducial model for the EKL scenario, which reproduces the observed eccentricity distribution of cold Jupiters from an initial Rayleigh distribution with a mean of 0.13. The librating near-resonant pairs, each of eight Earth masses, initially lie at around 0.1 au, with $\Delta \approx$ 0.01. In this study, the host star and the stellar companion have masses of 1 solar mass. The simulations are then run for 16 Myr to study the statistical evolution of TTV phases and the libration amplitude of resonant angles.

Compared to the same sample in the absence of a stellar companion, the stellar EKL tends to excite a higher eccentricity of a cold Jupiter at larger distances ($P_{\rm CJ}/P_{\rm SE1} > 50$) and thus brings a giant planet closer to the resonant pair. The close approach can induce non-secular interactions that break the resonant pair, or enhance secular interactions, thereby exciting TTV phases and the libration amplitude of resonant angles away from their initial geometric alignment.
We find that $\approx$25\% of the resonant pairs in our model are disrupted by a highly inclined stellar companion. Up to $\approx 20$\% of the pair members can attain the maximum TTV phase of $180^\circ$, compared to $\lesssim 12.5$\% without EKL. The libration amplitude of resonant angles is easier to excite than the TTV phases, because circulating TTV phases require the free eccentricity to be much larger than the forced eccentricity. $\approx$30-50\% of the EKL-perturbed pair members can reach circulating resonances. We also run the r0.13 model in the presence of a stellar companion for up to 100 Myr and find that $\sim 5$\% more of the resonant pairs are disrupted by large TTV phases, by extension, large free eccentricities, induced by stellar EKL during the first 16 Myr of our simulations. 

We also compare the same sample with an initial nearly circular orbit of a cold Jupiter. The EKL excitation of initial TTV phases and libration amplitudes is reduced by $\sim 20$\%.

Because the statistical studies cover a wide range of orbital distributions for a cold Jupiter and a stellar companion, a substantial fraction of resonant planet pairs in our parameter space are subject to weak to moderate EKL perturbations. Resonant pairs experiencing moderate EKL forcing, characterized by $\mathcal{D}/\Delta \sim 1$, develop asymmetric TTV phase distributions, i.e., they display a pronounced deviation from perfect anti-correlation. This effect is particularly evident for 2:1 resonant pairs, owing to their comparatively weaker dynamical coupling. 
We discuss that although the observational uncertainties in the Kepler TTV measurements do not preclude the presence of such asymmetric TTV phases, the Kepler systems with $0 < \Delta < 0.03$ generally exhibit more nearly symmetric TTV correlations than those produced in our numerical model. This discrepancy suggests that additional dynamical excitation mechanisms, beyond the EKL process considered in this work, may be operating in close-in resonant super-Earth systems.
Nonetheless, our statistical studies show a paucity of systems with maximum TTV phases in the range $\sim 90^\circ$–$180^\circ$, favoring small but non-zero TTV phases, in agreement with the trends inferred from the Kepler TTV data.

\begin{acknowledgments}
We thank the referee for the useful comments, which greatly improved this paper.
We are deeply grateful to Yoram Lithwick for illuminating discussions on TTV theories. 
P.-G.G. thanks Kangrou Guo, Hsinling Hu, Xiumin Huang, Yuji Matsumoto, and Hsi-Wei Yen for useful discussions.  The numerical work was
conducted on the high-performance computing facility at the
Institute of Astronomy and Astrophysics in Academia Sinica. P.-G.G. acknowledges support from the National Science and Technology Council in Taiwan (NSTC) under grant 114-2112-M-001-043. G. Li thanks the support of the Visiting Miller Professorship, funded by the Miller Institute at UC Berkeley.
\end{acknowledgments}

\appendix
\section{Resonant capture in the presence of a cold Jupiter: a numerical test}
\label{sec:capture}

Because an initial Rayleigh distribution of $e_{\rm CJ}$ is likely produced by scattering among giant planets after the gas disk is dissipated \citep{2025ApJ...980L..31W}, we introduce a CJ into the numerical integrations after turning off the dissipative forces responsible for planet migration in a gaseous disk. However, a CJ should have formed within a protoplanetary disk in order to accrete a substantial gaseous envelope. 
Jupiter-mass planets could gain orbital eccentricities up to the disk's aspect ratio $\lesssim 0.1$ through planet-disk tidal interactions \citep{2015ApJ...812...94D}, which may affect the outcome of resonance capture.

\begin{figure}[ht!]
\epsscale{0.7}
\plotone{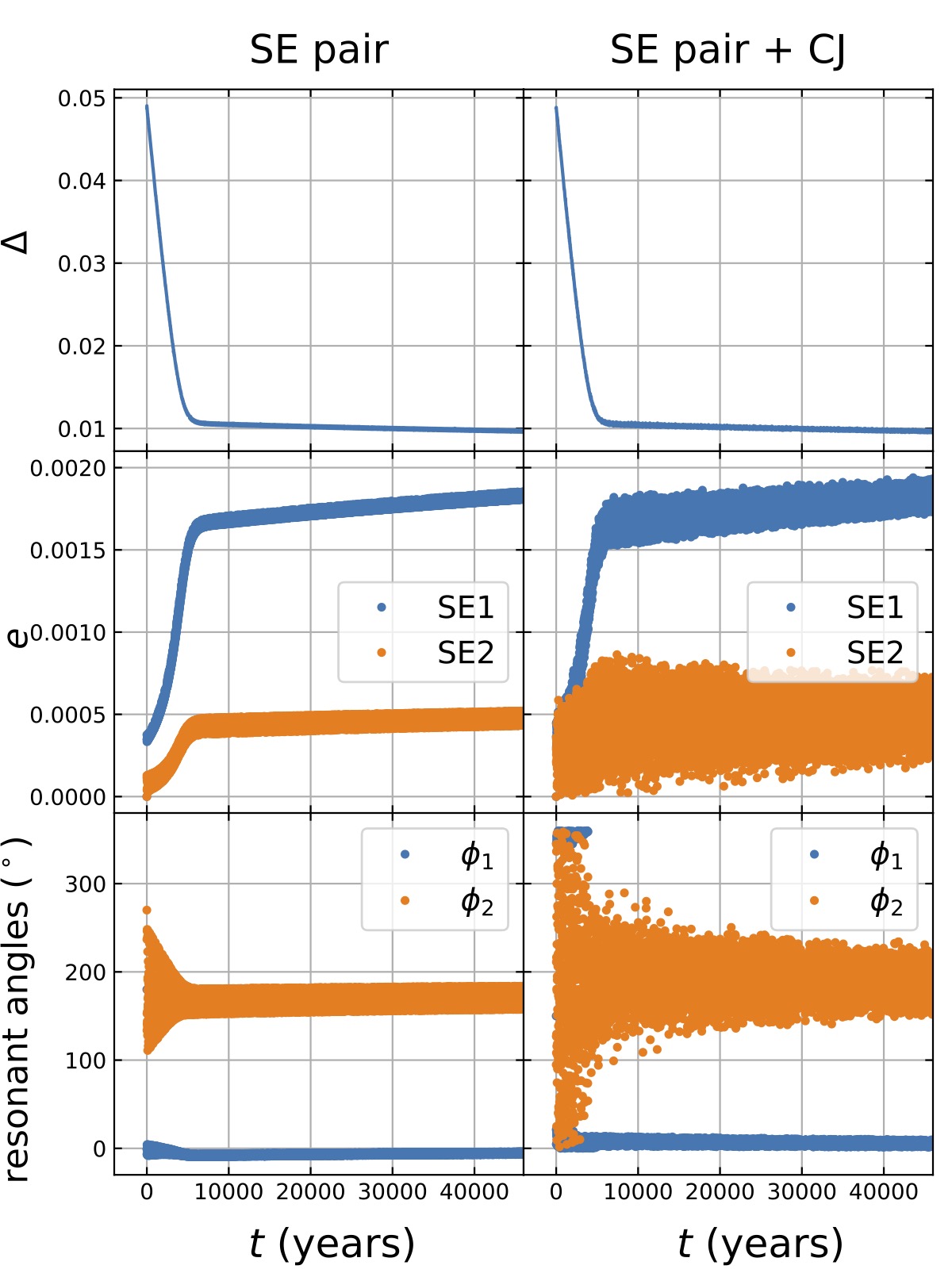}
\caption{Comparison of the 2:1 resonance capture in the absence (left panels) and presence (right panels) of a CJ of 1 $M_{\rm jup}$ on an eccentric orbit with $a_{\rm CJ}=0.5$ au and $e_{\rm CJ}=0.1$.
\label{fig:comp_init}}
\end{figure}

In Figure \ref{fig:comp_init}, we compare our initial setup of the 2:1 resonance capture without a CJ (left panels) to the same capture scenario including a 1 $M_{\rm jup}$ CJ (right panels). 
To accentuate any difference in MMR capture arising from the CJ’s perturbation, we choose $a_{\rm CJ}=0.5$ au (the inner edge of the CJ semimajor-axis range adopted in this work) and $e_{\rm CJ}=0.1$. There are no dissipative forces on the CJ's orbit. Even under this deliberately extreme numerical test, the SE pair remains robustly captured in the 2:1 MMR, as evidenced by $\Delta$ approaching $0.01$ (top panels), $e$ settling to near equilibrium values (middle panels), and the resonant angles evolving toward small libration amplitudes (bottom panels). We note, however, that the libration amplitudes of the resonant angles are larger when an eccentric CJ is present than in the case without a CJ. Thus, our adopted initial condition—starting from a dynamically cold, near-resonant pair—is a conservative choice for exploring subsequent evolution. The effect of a CJ on capture into the 3:2 MMR is analogous, so we omit showing those evolutions here for brevity.

Nonetheless, we stress that we use ad hoc dissipative forces to achieve resonance capture through convergent migration with strong eccentricity damping in our simulations, i.e., resonant replusion, which should not be interpreted as a physical process we intend to study but a standard numerical procedure to set up the initial condition with $\Delta \sim 0.01$. We do not explicitly model the formation of $\Delta \sim 0.01$ in this study (see Section \ref{sec:formation_Delta} for discussion on this issue).

\section{TTV phases of a near-resonant pair}
\label{sec:appendix}

\subsection{Model and Calculation}
\citet{2012ApJ...761..122L} formulated an analytical TTV theory for two coplanar planets near a first-order mean-motion resonance (MMR), under the assumption that the resonant angles are in the circulating regime. They showed that their theory remains valid provided the free eccentricity satisfies $e_{\rm free} \lesssim \Delta^2/\mu_{\rm SE}$. In addition, \citet{2023MNRAS.522.1914C} verified that the L12 model remains applicable to librating first-order MMRs when $\Delta/\mu_{\rm SE}^{2/3} \gtrsim 1$. 
In our sample of pairs that do not experience resonant disruption, we typically have $\Delta \approx 0.01$–0.02 and $\mu_{\rm SE} = 2.4 \times 10^{-5}$, yielding $\Delta / \mu_{\rm SE}^{2/3} \approx 12 > 1$.\footnote{It can also be shown that $\Delta / \mu_{\rm SE}^{2/3}$ is approximately equal to the parameter $\bar{\delta}$ introduced by \citet{1999ssd..book.....M} as the bifurcation criterion for a first-order MMR. A large positive value, $\bar{\delta} \approx 12$, corresponds to the resonant regime, analogous to Figure 8.10(a) of \citet{1999ssd..book.....M}, in which the resonant pairs possess only a single stable libration center in the phase space of the resonant Hamiltonian. Consequently, the MMRs in our study are far from exact commensurability and are sufficiently weak to permit multiple equilibrium points.} 
We have verified that the condition $e_{\rm free} \lesssim \Delta^2/\mu_{\rm SE} \sim 10$ is satisfied in our simulations, except in cases where resonance disruption (a non-secular process) occurs. 

Moreover, \citet{2016ApJ...823...72N} derived a TTV theory for librating MMRs and examined the $\Delta$–$\mu$ parameter space over which their TTV expressions reduce to the L12 model (see their Figure 10). They further noted that, for the L12 model to be applicable, the eccentricity of a resonant pair must satisfy $e \lesssim \sqrt{\Delta}$. 
We find that our adopted values of $\Delta$ and $\mu_{SE}$, together with the eccentricities measured in simulations of non-disruptive resonant pairs, lie within the parameter domain identified by Nesvorn\'y \& Vokrouhlick\'y in which the L12 framework remains valid for librating MMRs. For a more rigorous validation, we follow \citet{2023MNRAS.522.1914C} and extract the TTV signals of SE1 and SE2 directly from the O–C (Observed minus Calculated), where the observed transit times are obtained from the \textsc{Rebound} IAS15 integrator with high-cadence outputs, and the calculated transit times are derived from a linear ephemeris fit assuming a constant orbital period \citep[e.g.,][]{2018haex.bookE...7A}. Based on these diagnostics, we confirm that the L12 model is applicable to both librating and asymmetric first-order MMRs across the parameter space explored in our simulations. We therefore adopt the L12 formalism for TTV calculations in both librating and circulating resonant configurations.

Near a first-order MMR, the complex eccentricity of a planet, $z=e \exp(i \varpi)$, can be decomposed into free and forced components. 
The L12 model derived analytical expressions for the forced eccentricity $z_{\rm forced}$ and the TTV cycles $V_{1}=|V_{1}|\exp(i \Phi_{\rm TTV,1})$ for the inner pair member, and the same for $V_2$ for the outer planet, as functions of the free eccentricity $z_{\rm free}$.
Hence, we first obtain the free eccentricity $z_{\rm free}=z-z_{\rm forced}$, where $z$ is taken from simulations at time $t$. Once $z_{\rm free}=e_{\rm free} \exp(i \varpi_{\rm free})$ is known, we then calculate the TTV phase $\Phi_{\rm TTV}$:
\begin{eqnarray}
\tan( \Phi_{\rm TTV,1}) &=&
{  -( e_{\rm free,1} \sin \varpi_{\rm free,1} - |f_2/f_1| e_{\rm free,2} \sin \varpi_{\rm free,2} ) \over
{2\over 3}\Delta + \left( e_{\rm free,1} \cos \varpi_{\rm free,1} - |f_2/f_1| e_{\rm free,2} \cos \varpi_{\rm free,2} \right) } = {\Im(\mathcal{Z}_{\rm free})/|f_1| \over {2\over 3}\Delta - \Re(\mathcal{Z}_{\rm free})/|f_1|}, \label{eq:Phi1} \\
\tan (\Phi_{\rm TTV,2}) &=& 
{  -( e_{\rm free,1} \sin \varpi_{\rm free,1} - |f_2/f_1|  e_{\rm free,2} \sin \varpi_{\rm free,2} ) \over
{2\over 3} |f_2/f_1| \Delta + \left( e_{\rm free,1} \cos \varpi_{\rm free,1} - |f_2/f_1|  e_{\rm free,2} \cos \varpi_{\rm free,2} \right) } 
= {\Im(\mathcal{Z}_{\rm free})/|f_1| \over {2\over 3}|f_2/f_1|\Delta - \Re(\mathcal{Z}_{\rm free})/|f_1|}, \label{eq:Phi2} 
\end{eqnarray}
where ($f_1$, $f_2$) = ($-2.02$, 2.48) for a 3:2 MMR and ($-1.19$, 0.43) for a 2:1 MMR, and $\mathcal{Z}_{\rm free} \equiv -|f_1|z_{\rm free,1} +|f_2| z_{\rm free,2}$.  The TTV phases measure the phase difference between the double transit time and the nearest zero-crossing time of the TTV signals. Following the convention adopted by \citet{2023MNRAS.522.1914C}, the nearest descending zero-crossing is used for SE1, and the nearest ascending zero-crossing is used for SE2. However, if the double transit precedes the nearest zero-crossing, then $\Phi_{\rm TTV} > 0$; otherwise, $\Phi_{\rm TTV} < 0$ in this work.\footnote{The sign convention is opposite in \citet{2023MNRAS.522.1914C}, and the double transit is termed ``TC" (Transit Conjunction) in their work.} This sign convention also applies to asymmetric TTV signals.

\subsection{Dynamical Properties}
It is clear from Equations \ref{eq:Phi1} and \ref{eq:Phi2} that when $|\mathcal{Z}_{\rm free}|/|f_1| > (2/3)\Delta$ or $|\mathcal{Z}_{\rm free}|/|f_1| < (2/3)\Delta$, $\Phi_{\rm TTV,1} \approx \Phi_{\rm TTV,2} \sim 1$ (large) or $\sim 0$ (small), respectively.  
In the intermediate regime of modest TTV phases where $|\mathcal{Z}_{\rm free}|/|f_1| \sim (2/3)\Delta$,
$\Phi_{TTV,1}$ can differ significantly from $\Phi_{TTV,2}$. Specifically,
$\Phi_{\rm TTV,2}$ can be noticeably larger than $\Phi_{\rm TTV,1}$ in this intermediate regime for 2:1 MMRs \citep[e.g., see Figure 3 of][for details]{2026AJ....171..135W}. This is because $|f_2/f_1| \approx 0.36$, much smaller than 1 due to an indirect potential, leading to a larger departure from symmetric (i.e., from anti-correlated) TTV signals of the inner and outer pair members. 
In comparison, $\Phi_{\rm TTV,1}$ can be noticeably larger than $\Phi_{\rm TTV,2}$ in this intermediate regime of the 3:2 resonance, as also seen in our ensemble simulations (not shown). This arises because $|f_2/f_1| \approx 1.23 > 1$ \citep[e.g., see Figure 8 of][for details]{2012ApJ...761..122L}. However, $|f_2/f_1|$ for the 3:2 MMR is closer to unity than that for the 2:1, so the dynamical asymmetry is less prominent, as seen in our simulated statistical results.

In addition, $\varpi_1 \approx \varpi_2$ when the secular influence of the third body is significant.
Hence, $(3/2)|\mathcal{Z}_{\rm free}|/|f_1|$ is comparable to $\mathcal{D} \equiv |e_{\rm SE1}-|f_2/f_1|e_{\rm SE2}|$. Thus, the condition $(3/2)|\mathcal{Z}_{\rm free}|/|f_1| > \Delta$ simplifies to $\mathcal{D}/\Delta > 1$, which \citet{2023MNRAS.522.1914C} used as a proxy to identify a resonant pair with significant TTV phases. Although \citet{2023MNRAS.522.1914C} used $\varpi_1 \approx \varpi_2$ to obtain the expression $\mathcal{D}/\Delta$, this does not always happen in our ensemble simulations (e.g., see Section \ref{sec:example} and Figure \ref{fig:r013_EKLexample}). Nevertheless, we find that max($\mathcal{D}/\Delta$) remains a useful proxy for evaluating the dynamical activity of a resonant pair in our statistical study, regardless of the apsidal alignment.
When max($\mathcal{D}/\Delta) \sim 1$, this corresponds to the intermediate regime of the TTV theory, where asymmetric TTV signals are appreciable, and the resonant pair is dynamically more excited. When max($\mathcal{D}/\Delta) > 1$ (thus $e_{\rm free} > \Delta$), near-symmetric TTV signals occur because $e_{\rm free} \gg e_{\rm forced} \sim \mu_{\rm SE}/\Delta$, and thus the resonant pair is dynamically much hotter.

Although large TTV phase excursions require $e_{\mathrm{free}} \gg e_{\mathrm{forced}}$, the transition of the resonant angles from libration to circulation occurs already at the threshold $e_{\mathrm{free}} = e_{\mathrm{forced}}$ \citep{1999ssd..book.....M}. Therefore, the libration amplitudes of the resonant angles are more easily excited than the TTV phases.



\subsection{Comparison of the Kepler TTV phase distribution between different analyses}
\label{sec:KeplerTTV_comp}

Despite the significant observational uncertainties, the Kepler TTV phase distribution illustrated in the lower panel of Figure \ref{fig:TTVcomp}, which is based on \citet{2014ApJ...787...80H}, overall appears a more asymmetric TTV correlation than those from \citet[][see the upper left panel of their Figure 17]{2025AJ....169..323L} and from \citet[][see the left panel of their Figure 4]{2023MNRAS.522.1914C}, which are based on \citet{2015ApJS..217...16R}.

\begin{figure}[ht!]
    \centering
    \includegraphics[width=\linewidth]{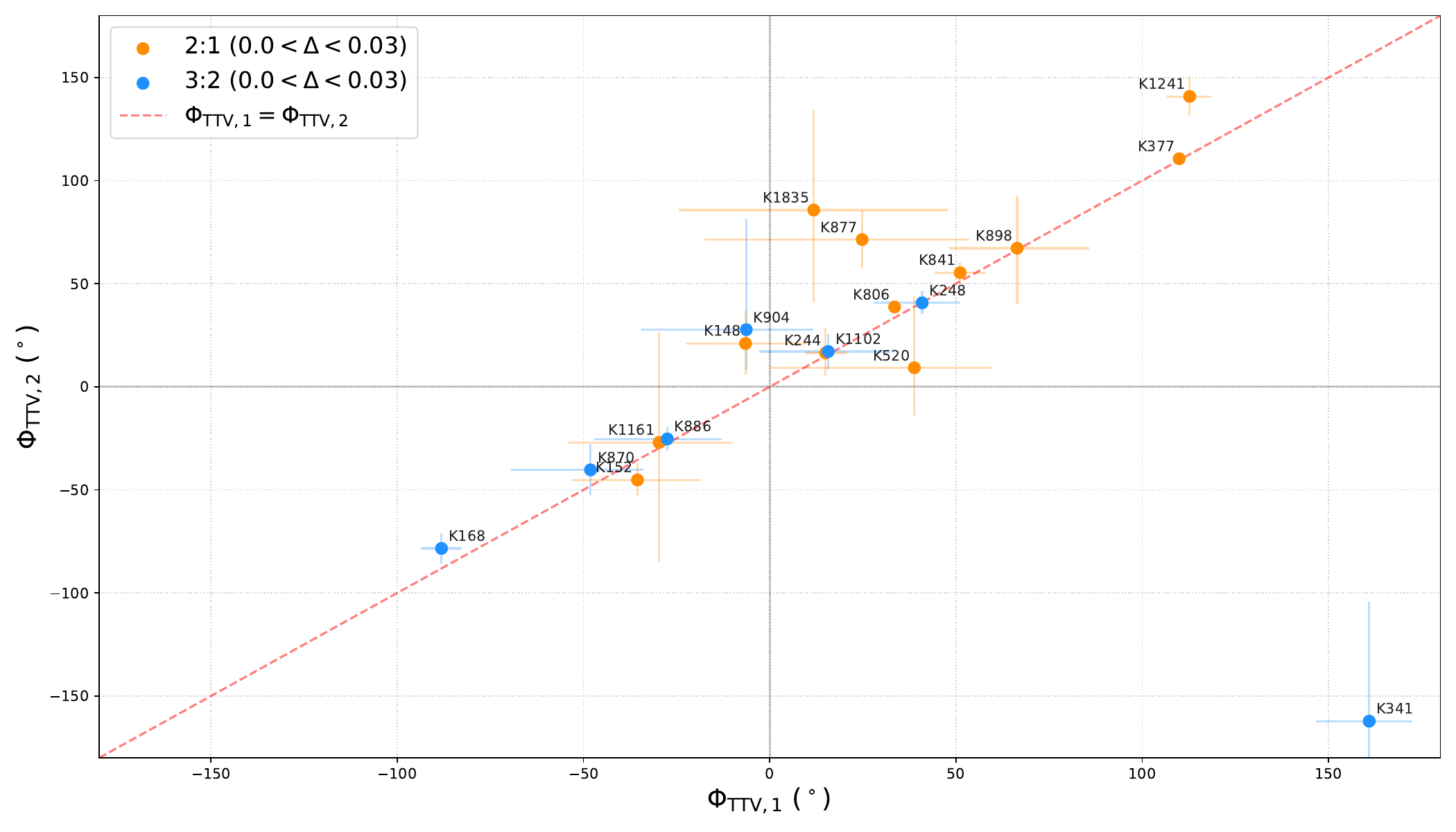}
\\ 
    \vspace{0.2cm} 
    \includegraphics[width=\linewidth]{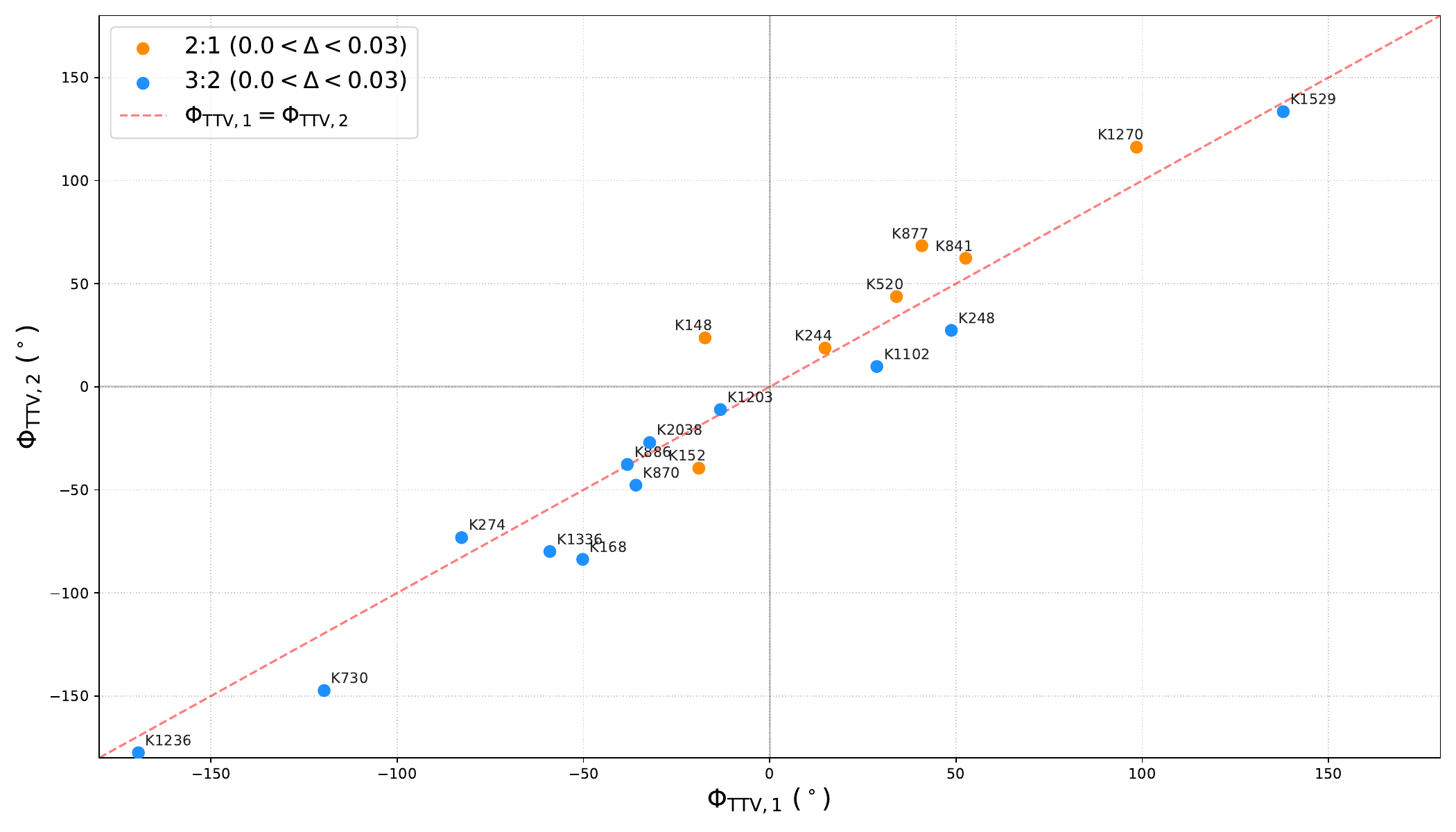}
    \caption{Comparison of the Kepler TTV phase distributions of near 2:1 and 3:2 MMRs from \citet{2014ApJ...787...80H} (upper panel) and \citet{2025AJ....169..323L} (lower panel). Kepler pairs with uncertainties greater than 60$^\circ$ have been
    removed in the upper panel. The interval of $\Delta$ is selected to range from 0 to 0.03 in these two distributions. The red dashed line corresponds to the exact symmetric correlation where $\Phi_{\rm TTV,1}=\Phi_{\rm TTV,2}$. The number following the letter 'K' labeled next to each data point indicates the corresponding KOI number of the host star. }
    \label{fig:KeplerTTV_comp}
\end{figure}

In Figure \ref{fig:KeplerTTV_comp}, we compare the distributions of Kepler TTV phases reported by \citet{2014ApJ...787...80H} (upper panel) and \citet{2025AJ....169..323L} (lower panel), both restricted to the same range of planetary separations, $0 < \Delta < 0.03$. Each data point is annotated with the KOI identifier of the corresponding planet-host star, thereby enabling a direct, object-by-object comparison between two studies. 
For consistency, the signs of the TTV phases from \citet{2025AJ....169..323L} have been inverted (i.e., positive values are mapped to negative and vice versa) to match the sign convention adopted by \citet{2024ApJ...971....5W} and in this work. The lower panel does not display uncertainty bars because phase uncertainties are not provided in \citet{2025AJ....169..323L}. The overlap between the two samples is partial: 10 KOIs are common to both analyses. 
The TTV phase distribution reported by \citet{2023MNRAS.522.1914C} closely resembles that from \citet{2025AJ....169..323L}.

Although different data analysis procedures were employed and there exist some discrepancies in KOI selection between the two samples, the resulting distributions become broadly similar in terms of MMR types (encoded by colors) and KOI identifications, in agreement with the expected symmetric TTV correlation about the red dashed line. KOI 1835, which appears in the top panel and whose measured correlation deviates from the exact symmetric relation at the level of 1$\sigma$, is not included in \citet{2025AJ....169..323L} because, in their analysis, the super-periods independently fitted to the TTV curves of the two planets in the pair differ significantly (refer to \dataset[the interactive dataset]{https://w.astro.berkeley.edu/~rixin/TTVs/MMR-TTV_v3.html} from \citet{2025AJ....169..323L}).

Note that additional asymmetric 2:1 mean-motion resonances are present over a wider range of $\Delta$ in the sample of \citet{2014ApJ...787...80H} (see Section \ref{sec:asymTTV} and the lower panel of Figure \ref{fig:TTVcomp}).\footnote{The additional Kepler systems are KOI 829 ($\Delta=0.03376$) and KOI 869 ($\Delta=0.03877$).} In contrast, \citet{2025AJ....169..323L} excludes systems with $\Delta > 0.03$ from their TTV analysis in order to maintain consistency with the observed $\Delta$ range adopted by \citet{2024AJ....168..239D}.

\section{Equations for timescales}
\label{sec:timescale}
The EKL cycle period 
is given by
\citep{2015MNRAS.452.3610A,2016ARA&A..54..441N}
\begin{eqnarray}
t_{\rm EKL} &=& {16\over 30 \pi}{m_* + m_{\rm CJ} + m_{\rm star2} \over m_{\rm star2}}{P^2_{\rm star2} \over P_{\rm CJ}} (1-e_{\rm star2}^2)^{3/2} \nonumber \\ 
&&\approx 15\,{\rm Myr} \left( {m_{\rm star2}  \over m_\odot} \right)^{-1}  \left( {m_*  \over m_\odot} \right)^{1/2}  \left( { a_{\rm star2} \over 1000\,{\rm au}} \right)^3 
 \left( {a_{\rm CJ} \over 5\,{\rm au}} \right)^{-3/2} (1-e^2_{\rm star2})^{3/2}.
\label{eq:t_EKL}
\end{eqnarray}
The precession timescale of the cold Jupiter due to the quadrupole gravitational potential of the resonant SE pair is \citep[e.g.,][]{1999ssd..book.....M}
\begin{eqnarray}
t_{\rm SE,J_2} &\approx& {4\over3} P_{\rm CJ}  {m_* a_{\rm CJ}^2 \over m_{\rm SE1} a_{\rm SE1}^2 + m_{\rm SE2} a_{\rm SE2}^2} \nonumber \\
&\approx & 775 \,{\rm Myr} \left( {m_*  \over m_\odot} \right)^{-1/2} \left( {a_{\rm CJ} \over 5\,{\rm au}} \right)^{7/2} \left( {a_{\rm SE} \over 0.1{\rm au}} \right)^{-2}  \left( {\mu_{\rm SE} \over 8m_\oplus / m_\odot} \right)^{-1},
\label{eq:t_J2}
\end{eqnarray}
where, for the numerical estimate, we have adopted $m_{\rm SE1}=m_{\rm SE2}$, $a_{\rm SE1}\approx a_{\rm SE2} \equiv a_{\rm SE}$, and defined the SE–host-star mass ratio as $\mu_{\rm SE} \equiv m_{\rm SE1}/m_*$.

\section{issue of disk tidal truncation against the location of CJs}
\label{sec:truncation}
We adopt the r0.13 model from \citet{2025ApJ...980L..31W} for the EKL oscillations of a cold Jupiter, which excites the TTV phases, or even disrupts a close-in near-resonant pair of super-Earths. For the initial conditions in this EKL model, $e_{\rm star2}$ is sampled uniformly from 0 to 1, and $a_{\rm star2}$ is drawn from a lognormal distribution from 50 to 1500 au. However, assuming that these initial conditions are inherited from the planet formation phase in a gaseous disk, the stellar companion could approach the disk closely enough to hinder the formation of a cold Jupiter.

To evaluate this formation issue, we take the population synthesis model from \citet{2026A&A...708A..38N}
for planet formation in an S-type binary, also known as the PAIRS project, named by the authors. Despite being restricted to coplanar orbits of S-type binaries, one simulation set of the PAIRS model (set A, see their Table 1) explored a wide range of initial conditions for disk, binary, and planet parameters, which nearly cover the ranges of $e_{\rm stars}$ and $a_{\rm star2}$ in our r0.13 model. Therefore, the model still provides a useful reference. The PAIRS model includes rapid core accretion via pebble accretion onto one planetary embryo. 

The model predicted that planets are able to grow up to $\sim 6 M_{\rm jup}$ at the distance $< 0.3 R_{\rm trunc}$ when the periastron of the binary is $> 20$ au. Here, $R_{\rm trunc}$ is the disk tidal truncation radius given by $R_{\rm Egg} \times (b e^c_{\rm star2} + 0.88 \mu^{0.01})$, where the stellar mass ratio $\mu=0.5$ and the Roche lobe radius $R_{\rm Egg} = 0.49 a_{\rm star2}/[0.6+ \ln(2)]$ for $m_* = m_{\rm star2}$ \citep{1983ApJ...268..368E}. $b$ and $c$ depend on the Reynolds number of the disk gas \citep[see Table C.1 of][]{2019A&A...628A..95M}. We adopt the largest Reynolds number ($\mathcal{R}=10^6$) tabulated in \citet{2019A&A...628A..95M}, and thus the weakest viscous disk torque against the tidal torque from the stellar companion, yielding the smallest $R_{\rm trunc}$ for a strict evaluation \citep{1994ApJ...421..651A}. This $\mathcal{R}$ value can correspond to the Shakura-Sunyaev viscosity parameter $\alpha=10^{-4}$ and the disk aspect ratio $h=0.1$.

\begin{figure}[htb!]
\plotone{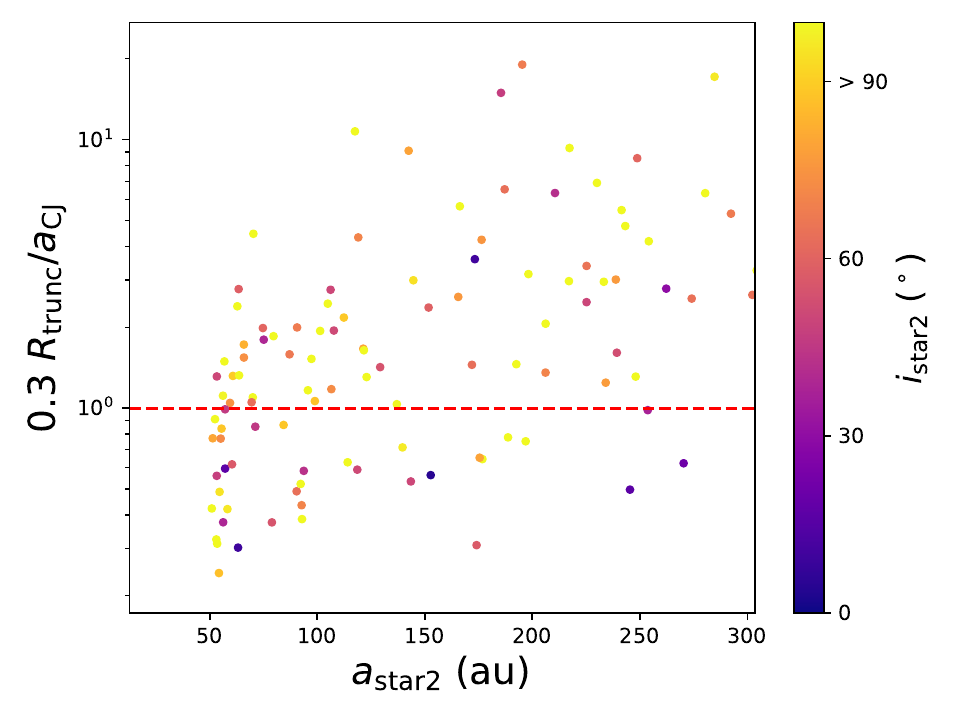}
\caption{The ratio of 0.3$R_{\rm trunc}$ to the initial values of $a_{\rm CJ}$ for the planetary systems in the r0.13 model. Only the cases with small binary separations, $a_{\rm star2} \lesssim 300$ au, are shown, as circumstellar disks are most affected in close binaries. The red dashed line indicates $0.3R_{\rm trunc} = a_{\rm CJ}$, below which a CJ is generally difficult to form via core accretion in the PAIRS model for a coplanar S-type binary.
The initial mutual inclination between each planetary system and its secondary stellar companion is indicated by the color scale. As $i_{\rm star2}$ increases, the tidal truncation torque exerted by the secondary star is reduced and thus $R_{\rm trunc}$ becomes larger than what is shown in the figure.
\label{fig:R_trunc}}
\end{figure}

Figure \ref{fig:R_trunc} presents the quantity $0.3 R_{\rm trunc}/a_{CJ}$ for the simulated systems in the r0.13 model, restricted to relatively close binaries with $a_{\rm star2} \lesssim 300$ au. In a substantial fraction of systems with $a_{\rm star2} \lesssim 100$ au, the CJ orbit lies beyond $0.3 R_{\rm trunc}$. However, as $a_{\rm star2}$ increases beyond 100 au, an increasing number of systems satisfy the formation condition $a_{\rm CJ} < 0.3 R_{\rm trunc}$. Moreover, we find that in most systems with $a_{\rm star2}>100$ au, the binary periastron distance exceeds 20 au (not shown). We therefore conclude that, in the r0.13 sample, CJs cannot generally form via pebble accretion in S-type binary systems with $a_{\rm star2}<100$ au in the PAIRS model.

Nevertheless, the truncation radius $R_{\rm trunc}$ can be larger when the binary orbit is inclined with respect to the disk midplane, because the tidal torque is reduced by a factor of $\cos^8 i_{\rm star2}$ \citep{2015ApJ...800...96L}. For reference, the mutual inclination $i_{\rm star2}$ is indicated by the color scale in Figure \ref{fig:R_trunc}.

\bibliography{ms_TTVEKL}{}
\bibliographystyle{ms_TTVEKL}



\end{document}